\documentclass[aps,preprint,onecolumn,superscriptaddress,showpacs,nofootinbibfloatfix,amsmath,amsfonts,amssymb]{revtex4-1}%
\usepackage{amsmath,amsfonts,amssymb,color}
\usepackage{amsthm}
\usepackage{leftidx}
\usepackage{graphicx}
\usepackage{xcolor}
\usepackage{dcolumn}
\usepackage{bm}
\usepackage{epstopdf}
\usepackage{epsfig}
\usepackage{environ}
\usepackage{pdfcomment}

\usepackage{multirow}
\usepackage{setspace}
\usepackage{color}

\usepackage{float}
\usepackage[T1]{fontenc}
\usepackage[latin9]{inputenc}
\usepackage{setspace}
\usepackage{esint}

\providecommand{\tabularnewline}{\\}

\begin{document}

%\preprint{APS/123-QED}

\title{Floquet topological superconductors with many Majorana edge modes:
	topological invariants, entanglement spectrum and bulk-edge correspondence}

\author{Hailing Wu}
\affiliation{%
	College of Physics and Optoelectronic Engineering, Ocean University of China, Qingdao, China 266100
}

\author{Shenlin Wu}
\affiliation{%
	College of Physics and Optoelectronic Engineering, Ocean University of China, Qingdao, China 266100
}

\author{Longwen Zhou}
\email{zhoulw13@u.nus.edu}
\affiliation{%
	College of Physics and Optoelectronic Engineering, Ocean University of China, Qingdao, China 266100
}
\affiliation{%
	Key Laboratory of Optics and Optoelectronics, Qingdao, China 266100
}
\affiliation{%
	Engineering Research Center of Advanced Marine Physical Instruments and Equipment of MOE, China 266100
}

\date{\today}

\begin{abstract}
One-dimensional Floquet topological superconductors possess two types
of degenerate Majorana edge modes at zero and $\pi$ quasienergies,
leaving more room for the design of boundary time crystals and quantum
computing schemes than their static counterparts. In this work, we
discover Floquet superconducting phases with large topological invariants
and arbitrarily many Majorana edge modes in periodically driven Kitaev
chains. Topological winding numbers defined for the Floquet operator
and Floquet entanglement Hamiltonian are found to generate consistent
predictions about the phase diagram, bulk-edge correspondence and
numbers of zero and $\pi$ Majorana edge modes of the system under
different driving protocols. The bipartite entanglement entropy further
shows non-analytic behaviors around the topological transition point
between different Floquet superconducting phases. These general features
are demonstrated by investigating the Kitaev chain with periodically
kicked pairing or hopping amplitudes. Our discovery reveals the rich
topological phases and many Majorana edge modes that could be brought
about by periodic driving fields in one-dimensional superconducting
systems. It further introduces a unified description for a class of
Floquet topological superconductors from their quasienergy bands and
entanglement properties.
\end{abstract}

%Use showkeys class option if keyword
\maketitle

\section{Introduction}\label{sec:Int}
Floquet topological phases of matter appear in systems with time-periodic
modulations. They have attracted great attention over the past decade
\cite{FloRev1,FloRev2,FloRev3,FloRev4,FloRev5}. The coupling of periodic
driving fields to a system could not only deform its band structures
and open topological gaps \cite{FloHF1,FloHF2,FloHF3,FloHF4,FloHF5,FloHF6},
but also generate new symmetry classifications \cite{FloCls1,FloCls2,FloCls3,FloCls4}
and anomalous Floquet phases with no static counterparts \cite{FloLTI1,FloLTI2,FloLTI3,FloLTI4,FloLTI5,FloLTI6,FloLTI7,FloLTI8,FloLTI9,FloLTI10,FloLTI11,FloLTI12,FloLTI13,FloLTI14,FloLTI15,FloLTI16,FloLTI17}.
The realization of Floquet topological matter in solid state and artificial
systems \cite{FloExp1,FloExp2,FloExp3,FloExp4,FloExp5,FloExp6,FloExp7,FloExp8,FloExp9,FloExp10,FloExp11,FloExp12}
further promotes their applications in ultrafast electronics \cite{FloRev3}
and quantum computation \cite{FloQC1,FloQC2,FloQC3,FloQC4,FloQC5}.

When applied to a pristine and noninteracting lattice model, periodic
driving fields could mainly generate three effects. At high frequencies,
the driving field could break the symmetry of the originally static
system and induce gap closing/reopening transitions in its spectrum,
especially around gapless points like Dirac cones \cite{FloHF1}.
This effect can usually be taken into account by implementing perturbative
expansions over the driving frequency \cite{FloHF3,FloHF4,FloHF5,FloHF6}.
When the driving amplitude and frequency of the field are comparable
with other energy scales of the system, more drastic changes could
appear in its spectral and topological properties. First, the Floquet
bands of both the bulk and edge states could develop crossings and
windings around the whole quasienergy Brillouin zone, yielding anomalous
Floquet topological metals \cite{FloLTI6,FloLTI13,FloLTI17} and insulators
\cite{FloLTI2,FloLTI5,FloLTI8} without any static analogies. Second,
spatially non-decaying, long-range couplings could be induced between
different lattice sites, yielding Floquet systems with large topological
invariants \cite{FloLTI1,FloLTI9}, many topological edge states \cite{FloLTI3,FloLTI4}
and multiple topological phase transitions \cite{FloLTI10,FloLTI11,FloLTI12}.
By designing suitable driving protocols, the last effect has been
utilized to obtain Floquet topological insulators with arbitrarily
large winding numbers and arbitrarily many dispersionless/chiral edge
and corner states \cite{FloLTI3,FloLTI9,FloLTI10,FloLTI11,FloLTI12,FloLTI14,FloLTI15}.
It has also been employed to realize Floquet Chern insulators with
large Chern numbers in both theory \cite{FloLTI7} and experiments
\cite{FloExp11}. However, much less is known regarding this effect
in Floquet superconducting systems \cite{FloTSC1,FloTSC2,FloTSC3,FloTSC4,FloTSC5}.
Specially, could we also generate Floquet superconductors with arbitrarily
large topological numbers and unboundedly many Majorana edge modes
with the help of driving-induced long-range couplings? A systematic
exploration of this problem could not only expand the usage of Floquet
engineering in superconducting setups, but also provide a rich source
of Majorana edge modes that might be used to realize more complicated
operations in boundary time crystals and Floquet quantum computing
\cite{FloQC3,FloQC4,FloQC5}.

In this work, we uncover the richness of phases and transitions in
one-dimensional (1D) Floquet topological superconductors. We begin
with an overview of the Ising-Kitaev chain \cite{KC} regarding its
symmetries, topological properties, Majorana edge modes, entanglement
spectrum (ES) and entanglement entropy (EE) in Sec.~\ref{sec:KC}.
Next, we introduce two representative models of Floquet Kitaev chains
by adding time-periodic kicks to the hopping or pairing amplitudes
of the system in Sec.~\ref{sec:PKKC}. Methods of characterizing the
topological and entanglement features of these Floquet superconducting
models in terms of their winding numbers, edge states, ES and EE will
also be discussed. In Sec.~\ref{sec:Res}, we systematically explore
the phases and transitions in our kicked Kitaev chains by investigating
the topological properties of their Floquet bands and Floquet entanglement
Hamiltonians. The emergence of Floquet superconducting phases with
large topological invariants and many Majorana zero/$\pi$ edge modes
will be revealed from the topological phase diagrams, quasienergy
spectrum and ES for each model. In Sec.~\ref{sec:Sum}, we summarize
our results and discuss potential future directions.

\section{Kitaev chain}\label{sec:KC}
In this section, we recap the Kitaev chain (KC) model, which describes a 1D superconducting wire with p-wave pairings \cite{KC}. We first introduce the model and obtain its bulk spectrum under the periodic boundary condition (PBC). We next introduce a winding number to characterize its topological phases and establish the topological phase diagram. This is followed by the presentation of the spectrum of KC under the open boundary condition (OBC) and the discussion of the correspondence between its bulk winding number and Majorana edge modes. Finally, in terms of the ES, EE and entanglement winding number, we characterize the topological properties of KC from the entanglement perspective. A collection of the quantities introduced in this section is summarized in Table \ref{tab:1}.

\begin{figure}
	\begin{centering}
		\includegraphics[scale=0.75]{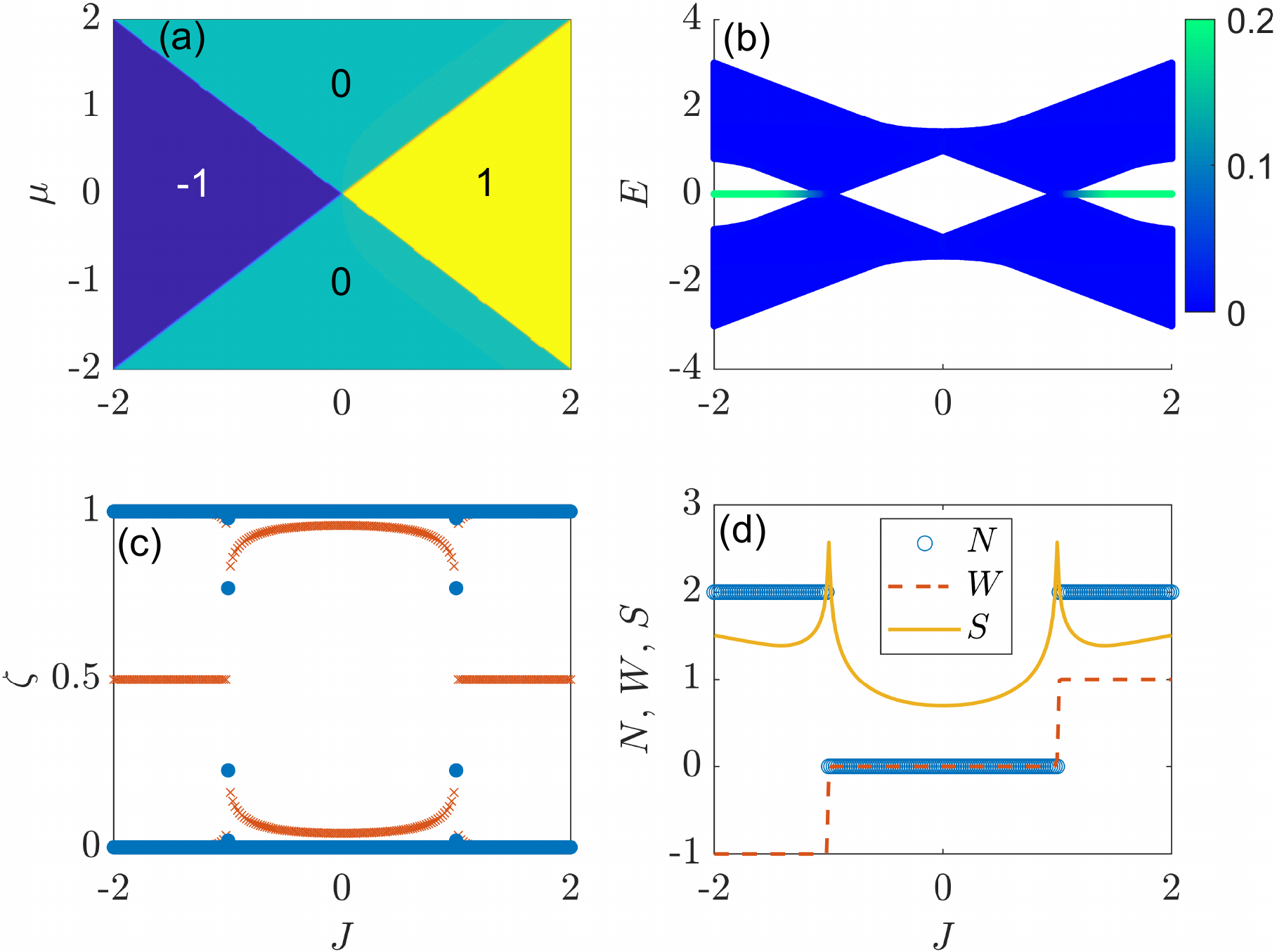}
		\par\end{centering}
	\caption{Phase diagram, spectrum, and entanglement properties of the KC. We set $\Delta=1$ for all panels and $\mu=1$ for panels (b), (c) and (d).
		(a) shows the winding number {[}Eq.~(\ref{eq:w}){]} versus hopping
		amplitude $J$ and chemical potential $\mu$. Each region with a uniform
		color corresponds to a topological phase with the value of $w$ highlighted
		therein. (b) shows the spectrum versus $J$ under the OBC.
		The color of each data point represents the inverse participation
		ratio (IPR) of the corresponding state \cite{Note1}. (c) shows the
		ES versus $J$, in which localized eigenmodes of the entanglement
		Hamiltonian are highlighted by red crosses. (d) shows the number $N$ of
		$\zeta=1/2$ entanglement eigenmodes, the winding number $W$ of the entanglement
		Hamiltonian and the EE $S$ versus $J$. \label{fig:KC}}
\end{figure}

The second quantized Hamiltonian of KC takes the form of
\begin{equation}
	\hat{H}=\frac{1}{2}\sum_{n}[\mu(\hat{c}_{n}^{\dagger}\hat{c}_{n}-1/2)+J\hat{c}_{n}^{\dagger}\hat{c}_{n+1}+\Delta\hat{c}_{n}\hat{c}_{n+1}+{\rm H.c.}].\label{eq:H}
\end{equation}
Here $c_{n}^{\dagger}$ $(c_{n})$ creates (annihilates) a spinless
fermion on the lattice site $n$. $\mu$ is the chemical potential.
$J$ is the nearest-neighbor hopping amplitude. $\Delta$ is the superconducting
pairing amplitude. For a chain of length $L$ and under the
PBC, we can perform the Fourier transformation
$\hat{c}_{n}=\frac{1}{\sqrt{L}}\sum_{k}e^{ikn}\hat{c}_{k}$ to find
the expression of $\hat{H}$ in momentum space as
\begin{equation}
	\hat{H}=\frac{1}{2}\sum_{k}\hat{\Xi}_{k}^{\dagger}H(k)\hat{\Xi}_{k},\label{eq:Hk}
\end{equation}
where $\hat{\Xi}_{k}^{\dagger}=\begin{pmatrix}\hat{c}_{k}^{\dagger} & \hat{c}_{-k}\end{pmatrix}$
is the Nambu spinor operator and the quasimomentum $k$ is defined
in the first Brillouin zone (BZ) $(-\pi,\pi]$. The Bloch Hamiltonian
\begin{equation}
	H(k)=\Delta\sin k\sigma_{y}+(\mu+J\cos k)\sigma_{z},\label{eq:hk}
\end{equation}
where $\sigma_{y}$ and $\sigma_{z}$ are Pauli matrices. The energy
dispersion of $H(k)$ takes the form
\begin{equation}
	\pm\varepsilon(k)=\pm\sqrt{\Delta^{2}\sin^{2}k+(\mu+J\cos k)^{2}},\label{eq:ek}
\end{equation}
which could become gapless at $k=0$ (BZ center) and $k=\pi$ (BZ
edge) if $\mu=-J$ and $\mu=J$, respectively. $H(k)$ possesses the
time-reversal symmetry ${\cal T}={\cal K}$ with ${\cal T}H(k){\cal T}^{-1}=H(-k)$,
the particle-hole symmetry ${\cal C}=\sigma_{x}{\cal K}$ with ${\cal C}H(k){\cal C}=-H(-k)$
and the chiral symmetry ${\cal S}=\sigma_{x}$ with ${\cal S}H(k){\cal S}=-H(k)$.
It thus belongs to the symmetry class BDI \cite{TPC1,TPC2}. Each of its topological
phases can be characterized by an integer-quantized winding number,
which is defined as \cite{KC2}
\begin{equation}
	w=\int_{-\pi}^{\pi}\frac{dk}{4\pi}{\rm Tr}[{\cal S}{\cal Q}(k)i\partial_{k}{\cal Q}(k)].\label{eq:w}
\end{equation}
Here ${\cal S}$ is the chiral symmetry operator and the projector
\begin{equation}
	{\cal Q}(k)=|\varepsilon^{+}(k)\rangle\langle\varepsilon^{+}(k)|-|\varepsilon^{-}(k)\rangle\langle\varepsilon^{-}(k)|.\label{eq:Q}
\end{equation}
$|\varepsilon^{+}(k)\rangle$ and $|\varepsilon^{-}(k)\rangle$ are
the eigenstates of $H(k)$ with energies $\varepsilon(k)$ and $-\varepsilon(k)$,
respectively. In Fig.~\ref{fig:KC}(a), we show the winding number
$w$ versus $J$ and $\mu$ at $\Delta=1$, which defines the topological
phase diagram of the Kitaev chain. We find $|w|=1$ for $|J|>|\mu|$
and $w=0$ for $|J|<|\mu|$. In the former case, the coupling between
Majorana modes in adjacent unit cells is stronger then their coupling
within each unit cell. The Kitaev chain then belongs to the topologically
nontrivial phase, supporting two Majorana zero modes at its two ends
under the OBC. In the latter case, the intracell
coupling between adjacent Majorana modes is stronger, and the Kitaev
chain belongs to the trivial phase without Majorana edge states.

To see the bulk-edge correspondence of the KC more clearly,
we consider its spectrum under the OBC. Applying the transformation
between two fermionic bases $\{\hat{c}_{n},\hat{c}_{n}^{\dagger}\}$ and
$\{\hat{f}_{j},\hat{f}_{j}^{\dagger}\}$ \cite{BdG}, i.e.,
\begin{equation}
	\hat{c}_{n}=\sum_{j=1}^{L}(u_{jn}\hat{f}_{j}+v_{jn}^{*}\hat{f}_{j}^{\dagger}),\qquad\hat{c}_{n}^{\dagger}=\sum_{j=1}^{L}(u_{jn}^{*}\hat{f}_{j}^{\dagger}+v_{jn}\hat{f}_{j}),\label{eq:BdG}
\end{equation}
we arrive at the following BdG self-consistent equations
\begin{alignat}{1}
	+\mu u_{n}+\frac{J}{2}(u_{n-1}+u_{n+1})+\frac{\Delta}{2}(v_{n-1}-v_{n+1}) & =Eu_{n},\label{eq:BdG1}\\
	-\mu v_{n}-\frac{J}{2}(v_{n-1}+v_{n+1})-\frac{\Delta}{2}(u_{n-1}-u_{n+1}) & =Ev_{n}.\label{eq:BdG2}
\end{alignat}
The\emph{ }complex coefficients \textbf{$u_{jn}$} and $v_{jn}$ satisfy
the normalization condition $\sum_{j=1}^{L}(|u_{jn}|^{2}+|v_{jn}|^{2})=1$.
The system Hamiltonian is assumed to be diagonal in the basis $\{\hat{f}_{j},\hat{f}_{j}^{\dagger}\}$.
Eqs.~(\ref{eq:BdG1}) and (\ref{eq:BdG2}) then follow by evaluating
the commutator $[\hat{H},\hat{c}_{n}]$ for $\hat{H}$ in its diagonal
form $\hat{H}=\sum_{j=1}^{L}E_{j}\hat{f}_{j}^{\dagger}\hat{f}_{j}$
and in the form of Eq.~(\ref{eq:H}) separately, and then letting
them to be equal \cite{BdG}. Note in passing that the redundant index
$j$ has been dropped in the final equations. The spectrum obtained
by solving the BdG equations (\ref{eq:BdG1}) and (\ref{eq:BdG2})
under the OBC is shown in Fig.~\ref{fig:KC}(b). We observe a pair
of degenerate edge modes at $E=0$ ($n_{0}=2$) in the topological
nontrivial regime ($|J|>|\mu|$) of the bulk with $|w|=1$ {[}see also
Fig.~\ref{fig:KC}(a){]}, which verifies the bulk-edge correspondence
$n_{0}=2|w|$ of the KC.

The topological nature of KC can also be revealed by its
bipartite ES and EE. The Eqs.~(\ref{eq:BdG1}) and (\ref{eq:BdG2})
can be equivalently generated by a synthetic Hamiltonian of spin-$1/2$
fermions
\begin{equation}
	\hat{{\cal H}}=\frac{1}{2}\sum_{n}[\mu\hat{\boldsymbol{{\bf c}}}_{n}^{\dagger}\sigma_{z}\hat{\boldsymbol{{\bf c}}}_{n}+\hat{\boldsymbol{{\bf c}}}_{n}^{\dagger}(J\sigma_{z}-i\Delta\sigma_{y})\hat{\boldsymbol{{\bf c}}}_{n+1}+{\rm H.c.}].\label{eq:HBdG}
\end{equation}
Here $\hat{\boldsymbol{{\bf c}}}_{n}^{\dagger}\equiv(\hat{c}_{n\uparrow}^{\dagger},\hat{c}_{n\downarrow}^{\dagger})$
and $\hat{c}_{n\sigma}^{\dagger}$ creates a fermion with spin $\sigma$
($=\uparrow,\downarrow$) on the lattice site $n$. Expressing a general
state as $|\psi\rangle=\sum_{n}(u_{n}\hat{c}_{n\uparrow}^{\dagger}+v_{n}\hat{c}_{n\downarrow}^{\dagger})|\emptyset\rangle$
and inserting it into the eigenvalue equation $\hat{{\cal H}}|\psi\rangle=E|\psi\rangle$,
we arrive at the Eqs.~(\ref{eq:BdG1}) and (\ref{eq:BdG2}), where
$|\emptyset\rangle$ represents the vacuum state with no particles.
At half-filling, decomposing the system under the PBC (with a ring
geometry) into two parts A and B of equal lengths $L/2$ and tracing
out all the degrees of freedom belonging to the subsystem B, we obtain
the reduced density matrix of the subsystem A as $\hat{\rho}_{{\rm A}}=\frac{1}{Z}e^{-\hat{H}_{{\rm A}}}$.
The eigenspectrum $\{\xi^{j}|j=1,...,L\}$ of entanglement Hamiltonian
$\hat{H}_{{\rm A}}$ defines the ES, and the von Neumann EE is given
by $S=-{\rm Tr}(\hat{\rho}_{{\rm A}}\ln\hat{\rho}_{{\rm A}})$. According
to Ref.~\cite{PeschelRev}, both the ES and EE can be expressed in
terms of the eigenvalues $\{\zeta^{j}|j=1,...,L\}$ of the system's
single particle correlation matrix $C_{mn}=\langle\hat{\boldsymbol{{\bf c}}}_{m}^{\dagger}\hat{\boldsymbol{{\bf c}}}_{n}\rangle$
($m,n\in{\rm A}$) as
\begin{equation}
	\xi^{j}=\ln(1/\zeta^{j}-1),\qquad\zeta^{j}\in[0,1],\label{eq:ES}
\end{equation}
\begin{equation}
	S=-\sum_{j=1}^{L}[\zeta^{j}\ln\zeta^{j}+(1-\zeta^{j})\ln(1-\zeta^{j})].\label{eq:EE}
\end{equation}
Due to the one-to-one correspondence between $\xi^{j}$ and $\zeta^{j}$,
we will also refer to the set $\{\zeta^{j}|j=1,...,L\}$ as the
ES. In Fig.~\ref{fig:KC}(c), we show the ES of the Kitaev chain.
Two entanglement eigenmodes at $\zeta=1/2$ {[}counted by the $N=2$
in Fig.~\ref{fig:KC}(d){]} are found only in the topological nontrivial
regime ($|J|>|\mu|$). Each of them is localized around the spatial
entanglement cuts between the two subsystems and yields a largest
contribution $\Delta S=\ln2$ to the EE. Discontinuous changes in
the number of these maximally entangled eigenmodes are observed at
$J=\pm\mu$, i.e., the topological transition points predicted by
the winding number $w$ of the Kitaev chain. They are also signified
by the non-analytic cusps in the EE of Fig.~\ref{fig:KC}(d). We can
further establish a bulk-edge correspondence for the entanglement
Hamiltonian $\hat{H}_{{\rm A}}$ by introducing a real-space winding
number for the subsystem A \cite{FloESEE}
\begin{equation}
	W=-\frac{1}{L'}{\rm Tr}'({\cal S}_{{\rm A}}Q[Q,N_{{\rm A}}]),\label{eq:W}
\end{equation}
where ${\cal S}_{{\rm A}}$ and $N_{{\rm A}}$ are the chiral symmetry
operator ($=\sigma_{x}$ for the Kitaev chain) and position operator
of the subsystem A. The trace is taken over the bulk region of A with
the size $L'=L_{{\rm A}}-2L_{{\rm E}}$, where $L_{{\rm A}}=L/2$
is the length of A and $L_{{\rm E}}$ counts the number of sites in
its left and right edge regions. The projector $Q\equiv\sum_{j}[\Theta(\zeta^{j}-1/2)-\Theta(1/2-\zeta^{j}|)]|\phi^{j}\rangle\langle\phi^{j}|$,
where the step function $\Theta(\zeta)=1$ ($=0$) if $\zeta>0$ ($\zeta<0$).
$\{\zeta^{j}\}$ and $\{|\phi^{j}\rangle\}$ are the eigenvalues and
eigenvectors of the correlation matrix $C$. In Fig.~\ref{fig:KC}(d),
we find $W=\pm1$ ($W=0$) in the topologically nontrivial (trivial)
region $|J|>|\mu|$ ($|J|<|\mu|$), and $W$ takes a unit jump when
the system undergoes a topological transition at $|J|=|\mu|$. Moreover,
we can identify an entanglement bulk-edge correspondence between $W$
and the number of $\zeta=1/2$ eigenmodes in the ES, i.e., $N=2|W|$.
Finally, we realize the relationship between the spectral and entanglement
topologies of the system, i.e., the edge modes $n_{0}=N$ and the
winding numbers $w=W$. Therefore, the ES, EE and entanglement winding
number $W$ can be used to completely characterize the topological
phases and transitions in the Kitaev chain. Note in passing that so
long as $\Delta\neq0$, the superconducting pairing strength does
not affect the phase diagram and topological transitions between different
superconducting phases in the KC. We will see that this
is not the case for our Floquet settings considered in Sec.~\ref{sec:Res}.

\begin{table}
	\begin{centering}
		\begin{tabular}{|c|c|}
			\hline 
			Quantity & Static Kitaev chain {[}Eq.~(\ref{eq:H}){]}\tabularnewline
			\hline 
			\hline 
			Winding number & $w=\int_{-\pi}^{\pi}\frac{dk}{4\pi}{\rm Tr}[{\cal S}{\cal Q}(k)i\partial_{k}{\cal Q}(k)]$\tabularnewline
			\hline 
			Entanglement spectrum & $\xi^{j}=\ln(1/\zeta^{j}-1)$\tabularnewline
			\hline 
			Entanglement entropy & $S=-\sum_{j}[\zeta^{j}\ln\zeta^{j}+(1-\zeta^{j})\ln(1-\zeta^{j})]$\tabularnewline
			\hline 
			Entanglement winding number & $W=-\frac{1}{L'}{\rm Tr}'({\cal S}_{A}Q[Q,N_{A}])$\tabularnewline
			\hline 
		\end{tabular}
		\par\end{centering}
	\caption{Various quantities and their definitions for the static KC. The ${\cal Q}(k)$ is defined in Eq.~(\ref{eq:Q}). The set $\{\zeta^{j}\}$
			includes all the eigenvalues of the correlation matrix $C_{mn}=\langle\hat{{\bf c}}_{m}^{\dagger}\hat{{\bf c}}_{n}\rangle$
			restricted to the subsystem A. The meanings of $L'$, ${\rm Tr}'$,
			${\cal S}_{A}$, $N_{A}$ and $Q$ are discussed below Eq.~(\ref{eq:W}).
		\label{tab:1}}
\end{table}

\section{Periodically kicked Kitaev chain\label{sec:PKKC}}

In this section, we introduce the models of Floquet topological superconductors that will be studied in this work. After presentating the Hamiltonians and Floquet operators of these models, we discuss how to characterize their quasienergy spectra and topological properties under different boundary conditions. We further introduce the definitions of ES, EE and real-space winding number for the Floquet entanglement Hamiltonian, which allow us to reveal the topological properties and bulk-edge correspondence of Floquet topological superconductors from the entanglement viewpoint. A set of key quantities introduced in this section is summarized in Table \ref{tab:2}.

We now add time-periodic modulations to the KC in Eq.~(\ref{eq:H}). We
consider adding $\delta$-kicks periodically to the pairing or hopping
amplitudes. The resulting model is thus described by the
Hamiltonian
\begin{equation}
	\hat{H}_{{\rm I}}(t)=\frac{1}{2}\sum_{n}[\mu(\hat{c}_{n}^{\dagger}\hat{c}_{n}-1/2)+J\hat{c}_{n}^{\dagger}\hat{c}_{n+1}+\Delta\delta(t)\hat{c}_{n}\hat{c}_{n+1}+{\rm H.c.}],\label{eq:H1}
\end{equation}
or
\begin{equation}
	\hat{H}_{{\rm II}}(t)=\frac{1}{2}\sum_{n}[\mu(\hat{c}_{n}^{\dagger}\hat{c}_{n}-1/2)+J\delta(t)\hat{c}_{n}^{\dagger}\hat{c}_{n+1}+\Delta\hat{c}_{n}\hat{c}_{n+1}+{\rm H.c.}],\label{eq:H2}
\end{equation}
where $\delta(t)\equiv\sum_{\ell\in\mathbb{Z}}\delta(t-\ell)$ and
we have assumed the driving period $T=1$. We will refer to these
periodically kicked Kitaev chains as PKKC1 {[}$\hat{H}_{{\rm I}}(t)${]}
and PKKC2 {[}$\hat{H}_{{\rm II}}(t)${]} for brevity. Their corresponding
Floquet operators, which describe the time evolution of the system
over a complete driving period (e.g., from $t=\ell+0^{-}$ to $t=\ell+1+0^{-}$)
are given by $\hat{U}_{s}=\hat{T}e^{-i\int_{0}^{1}\hat{H}_{s}(t)dt}$
for $s={\rm I,II}$, where $\hat{T}$ performs the time ordering.
They take the explicit forms
\begin{equation}
	\hat{U}_{{\rm I}}=e^{-\frac{i}{2}\sum_{n}[\mu(\hat{c}_{n}^{\dagger}\hat{c}_{n}-1/2)+J\hat{c}_{n}^{\dagger}\hat{c}_{n+1}+{\rm H.c.}]}e^{-\frac{i}{2}\sum_{n}\Delta(\hat{c}_{n}\hat{c}_{n+1}+{\rm H.c.})},\label{eq:U1}
\end{equation}
\begin{equation}
	\hat{U}_{{\rm II}}=e^{-\frac{i}{2}\sum_{n}[\mu(\hat{c}_{n}^{\dagger}\hat{c}_{n}-1/2)+\Delta\hat{c}_{n}\hat{c}_{n+1}+{\rm H.c.}]}e^{-\frac{i}{2}\sum_{n}J(\hat{c}_{n}^{\dagger}\hat{c}_{n+1}+{\rm H.c.})}.\label{eq:U2}
\end{equation}
Under the PBC, we can again perform Fourier transformations from position
to momentum representations and obtain 
\begin{equation}
	\hat{H}_{s}(t)=\frac{1}{2}\sum_{k}\hat{\Xi}_{k}^{\dagger}H_{s}(k,t)\hat{\Xi}_{k}\label{eq:H12k}
\end{equation}
for $s={\rm I,II}$, where
\begin{equation}
	H_{{\rm I}}(k,t)=\Delta\delta(t)\sin k\sigma_{y}+(\mu+J\cos k)\sigma_{z},\label{eq:h1k}
\end{equation}
\begin{equation}
	H_{{\rm II}}(k,t)=\Delta\sin k\sigma_{y}+[\mu+J\delta(t)\cos k]\sigma_{z}.\label{eq:h2k}
\end{equation}
The corresponding Floquet operators in the Nambu basis $\hat{\Xi}_{k}^{\dagger}=\begin{pmatrix}\hat{c}_{k}^{\dagger} & \hat{c}_{-k}\end{pmatrix}$
are thus given by
\begin{equation}
	U_{{\rm I}}(k)=e^{-i(\mu+J\cos k)\sigma_{z}}e^{-i\Delta\sin k\sigma_{y}},\label{eq:U1k}
\end{equation}
\begin{equation}
	U_{{\rm II}}(k)=e^{-i(\Delta\sin k\sigma_{y}+\mu\sigma_{z})}e^{-iJ\cos k\sigma_{z}}.\label{eq:U2k}
\end{equation}
We see that both of them can be expressed in the following piecewise form
\begin{equation}
	{\cal U}(k)=e^{-ih_{b}(k)}e^{-ih_{a}(k)},\label{eq:Uk}
\end{equation}
which does not possess any apparent symmetries. To determine the symmetry
class of a 1D Floquet system, we can first transform it into a pair
of symmetric time frames \cite{STM1,STM2,STM3}. For the ${\cal U}(k)$ in Eq.~(\ref{eq:Uk}),
these time frames can be found by splitting the time duration of its
kick or free evolution part by half. The resulting Floquet operators
in these time frames are
\begin{equation}
	{\cal U}_{1}(k)=e^{-\frac{i}{2}h_{a}(k)}e^{-ih_{b}(k)}e^{-\frac{i}{2}h_{a}(k)},\label{eq:UIk}
\end{equation}
\begin{equation}
	{\cal U}_{2}(k)=e^{-\frac{i}{2}h_{b}(k)}e^{-ih_{a}(k)}e^{-\frac{i}{2}h_{b}(k)}.\label{eq:UIIk}
\end{equation}
It is clear that ${\cal U}(k)$ and ${\cal U}_{1,2}(k)$ are related
by unitary transformations, which implies that they share the same
quasienergy spectrum. In the symmetric time frames, one can verify
that ${\cal U}_{\alpha}(k)$ ($\alpha=1,2$) possesses the chiral
symmetry ${\cal S}=\sigma_{x}$ with ${\cal S}{\cal U}_{\alpha}(k){\cal S}={\cal U}_{\alpha}^{\dagger}(k)$,
the time-reversal symmetry ${\cal T}={\cal K}$ with ${\cal T}{\cal U}_{\alpha}(k){\cal T}^{-1}={\cal U}_{\alpha}^{\dagger}(-k)$,
and the particle-hole symmetry ${\cal C}=\sigma_{x}{\cal K}$ with
${\cal C}{\cal U}_{\alpha}(k){\cal C}^{-1}={\cal U}_{\alpha}(-k)$
for both the PKKC1 and PKKC2. Therefore, the system described by ${\cal U}(k)$
also belongs to the BDI symmetry class \cite{STM1,STM2,STM3}. It further possesses the same
set of symmetries as the original KC. Each Floquet topological
phase of the PKKC1 or PKKC2 can then be characterized by a pair of
integer topological winding numbers $(w_{0},w_{\pi})\in\mathbb{Z}\times\mathbb{Z}$.

For our ${\cal U}(k)$, the $(w_{0},w_{\pi})$ can be obtained from
the winding numbers of ${\cal U}_{1}(k)$ and ${\cal U}_{2}(k)$ in
the symmetric time frames. Referring to the Eq.~(\ref{eq:w}), we
define
\begin{equation}
	w_{\alpha}=\int_{-\pi}^{\pi}\frac{dk}{4\pi}{\rm Tr}[{\cal S}{\cal Q}_{\alpha}(k)i\partial_{k}{\cal Q}_{\alpha}(k)],\qquad\alpha=1,2.\label{eq:w12}
\end{equation}
Here ${\cal S}=\sigma_{x}$ is the chiral symmetry operator and the
projector
\begin{equation}
	{\cal Q}_{\alpha}(k)=|\psi_{\alpha}^{+}(k)\rangle\langle\psi_{\alpha}^{+}(k)|-|\psi_{\alpha}^{-}(k)\rangle\langle\psi_{\alpha}^{-}(k)|.\label{eq:Q12}
\end{equation}
$|\psi_{\alpha}^{+}(k)\rangle$ and $|\psi_{\alpha}^{-}(k)\rangle$
are the eigenstates of ${\cal U}_{\alpha}(k)$ ($\alpha=1,2$) with
the quasienergies $E(k)$ and $-E(k)$, respectively. They are solutions
of the Floquet eigenvalue equation ${\cal U}_{\alpha}(k)|\psi_{\alpha}^{\pm}(k)\rangle=e^{-i(\pm)E(k)}|\psi_{\alpha}^{\pm}(k)\rangle$.
With these $w_{\alpha}$ ($\alpha=1,2$), we can express the topological
invariants of ${\cal U}(k)$ as
\begin{equation}
	w_{0}=\frac{w_{1}+w_{2}}{2},\qquad w_{\pi}=\frac{w_{1}-w_{2}}{2}.\label{eq:w0p}
\end{equation}
As will be seen shortly, the $(w_{0},w_{\pi})$ provide us with
complete topological characterizations for the Floquet superconducting
phases in both the PKKC1 {[}Eq.~(\ref{eq:U1}){]} and PKKC2 
{[}Eq.~(\ref{eq:U2}){]}.

Under the OBC, we can identify the presence of Floquet Majorana edge
modes from the quasienergy spectra and eigenstates of $\hat{U}_{{\rm I}}$
and $\hat{U}_{{\rm II}}$ for the two kicking protocols. Starting
with the Eq.~(\ref{eq:H1}) or (\ref{eq:H2}) and performing the
BdG transformation as presented in Eq.~(\ref{eq:BdG}), we can also
obtain the BdG self-consistent equations for the kicking systems.
For the $\hat{H}_{{\rm I}}(t)$ {[}Eq.~(\ref{eq:H1}){]} they are
given by
\begin{alignat}{1}
	+\mu u_{n}+\frac{J}{2}(u_{n-1}+u_{n+1})+\frac{\Delta\delta(t)}{2}(v_{n-1}-v_{n+1}) & ={\cal E}u_{n},\label{eq:BdG3}\\
	-\mu v_{n}-\frac{J}{2}(v_{n-1}+v_{n+1})-\frac{\Delta\delta(t)}{2}(u_{n-1}-u_{n+1}) & ={\cal E}v_{n}.\label{eq:BdG4}
\end{alignat}
For the $\hat{H}_{{\rm II}}(t)$ {[}Eq.~(\ref{eq:H2}){]} these equations
are
\begin{alignat}{1}
	+\mu u_{n}+\frac{J\delta(t)}{2}(u_{n-1}+u_{n+1})+\frac{\Delta}{2}(v_{n-1}-v_{n+1}) & ={\cal E}u_{n},\label{eq:BdG5}\\
	-\mu v_{n}-\frac{J\delta(t)}{2}(v_{n-1}+v_{n+1})-\frac{\Delta}{2}(u_{n-1}-u_{n+1}) & ={\cal E}v_{n}.\label{eq:BdG6}
\end{alignat}
Here ${\cal E}={\cal E}(t)$ denotes the instantaneous eigenvalue of the corresponding set of equations. Further derivation details of  Eqs.~(\ref{eq:BdG3})--(\ref{eq:BdG6}) can be found in the Appendix \ref{sec:UBdG}.
Integrating over a driving period, the Floquet operators of the systems
under both kicking protocols can be expressed in the unified form of
\begin{equation}
	\mathsf{U}=e^{-i\mathsf{H}_{b}}e^{-i\mathsf{H}_{a}}.\label{eq:UBdG}
\end{equation}
The eigensystem of $\mathsf{U}$ can then be obtained by solving the
eigenvalue equation $\mathsf{U}|\psi\rangle=e^{-iE}|\psi\rangle$
in the basis $(u_{1},v_{1},...,u_{n},v_{n},...,u_{L},v_{L})^{{\rm \top}}$.
A Floquet Majorana zero ($\pi$) mode refers to an eigenstate of $\mathsf{U}$
with the quasienergy $E=0$ ($E=\pm\pi$) and a localized profile
around the edges of the chain. When the system undergoes a topological
phase transition accompanied by the closing of its quasienergy gap
at $E=0$ ($E=\pm\pi$), we expect to see a quantized change in the number
$n_{0}$ ($n_{\pi}$) of Floquet Majorana zero ($\pi$) modes. As
will be seen in the next section, the winding numbers $w_{0}$ and
$w_{\pi}$ in Eq.~(\ref{eq:w0p}) could characterize the gap-closing/reopening
transitions in the Floquet spectra and the numbers of Majorana edge
modes at zero and $\pi$ quasienergies.

Similar to the static KC, we can also extract the topological
properties and bulk-edge correspondence of the Floquet KC
from its entanglement characteristics. One can first view the BdG
self-consistent equations as generated by synthetic Hamiltonians of
spin-$1/2$ fermions. For our PKKC1, the effective Hamiltonian that
could generate the Eqs.~(\ref{eq:BdG3}) and (\ref{eq:BdG4}) reads
\begin{equation}
	\hat{{\cal H}}_{{\rm I}}(t)=\frac{1}{2}\sum_{n}[\mu\hat{\boldsymbol{{\bf c}}}_{n}^{\dagger}\sigma_{z}\hat{\boldsymbol{{\bf c}}}_{n}+\hat{\boldsymbol{{\bf c}}}_{n}^{\dagger}(J\sigma_{z}-i\Delta\delta(t)\sigma_{y})\hat{\boldsymbol{{\bf c}}}_{n+1}+{\rm H.c.}].\label{eq:HBdG1}
\end{equation}
For the PKKC2, such a Hamiltonian is found to be {[}see Eqs.~(\ref{eq:BdG5})
and (\ref{eq:BdG6}){]}
\begin{equation}
	\hat{{\cal H}}_{{\rm II}}(t)=\frac{1}{2}\sum_{n}[\mu\hat{\boldsymbol{{\bf c}}}_{n}^{\dagger}\sigma_{z}\hat{\boldsymbol{{\bf c}}}_{n}+\hat{\boldsymbol{{\bf c}}}_{n}^{\dagger}(J\delta(t)\sigma_{z}-i\Delta\sigma_{y})\hat{\boldsymbol{{\bf c}}}_{n+1}+{\rm H.c.}].\label{eq:HBdG2}
\end{equation}
Integrating the corresponding Schrödinger equations over a driving
period, both the Floquet operators of $\hat{{\cal H}}_{{\rm I}}(t)$
and $\hat{{\cal H}}_{{\rm II}}(t)$ can be written in the piecewise
form of
\begin{equation}
	\hat{{\cal U}}=e^{-i\hat{{\cal H}}_{b}}e^{-i\hat{{\cal H}}_{a}},\label{eq:UESEE}
\end{equation}
which can be further expressed in two symmetric time frames \cite{STM1,STM2,STM3} as
\begin{equation}
	\hat{{\cal U}}_{1}=e^{-\frac{i}{2}\hat{{\cal H}}_{a}}e^{-i\hat{{\cal H}}_{b}}e^{-\frac{i}{2}\hat{{\cal H}}_{a}},\label{eq:UI}
\end{equation}
\begin{equation}
	\hat{{\cal U}}_{2}=e^{-\frac{i}{2}\hat{{\cal H}}_{b}}e^{-i\hat{{\cal H}}_{a}}e^{-\frac{i}{2}\hat{{\cal H}}_{b}}.\label{eq:UII}
\end{equation}
The bipartite ES and EE of the system can then be obtained following
the recipe of Ref.~\cite{FloESEE} (see also Ref.~\cite{FloESEE2}).
In short, we first divide the whole lattice of PKKC1 or PKKC2 under
the PBC into two half-chains A and B of equal lengths. Next, we assume
that initially the system is at half-filling, i.e., all the Floquet
eigenstates of $\hat{{\cal U}}_{\alpha}$ with quasienergies $E\in[-\pi,0)$
are filled ($\alpha=1,2$). Tracing out all the degrees of freedom
belonging to the subsystem B, we obtain the reduced density matrix
of subsystem A in the time frame $\alpha$ as $\hat{\rho}_{{\rm A}}^{\alpha}=\frac{1}{Z}e^{-\hat{H}_{{\rm A}}^{\alpha}}$.
Here $\hat{H}_{{\rm A}}^{\alpha}$ is the Floquet entanglement Hamiltonian
of subsystem A, whose eigenvalues define the ES in the $\alpha$th
time frame. For either the PKKC1 or PKKC2 and in the time frame $\alpha$,
the ES $\{\xi_{\alpha}^{j}|j=1,...,L\}$ and EE $S_{\alpha}$ are
given by \cite{FloESEE}
\begin{equation}
	\xi_{\alpha}^{j}=\ln(1/\zeta_{\alpha}^{j}-1),\label{eq:ES12}
\end{equation}
\begin{equation}
	S_{\alpha}=-\sum_{j=1}^{L}[\zeta_{\alpha}^{j}\ln\zeta_{\alpha}^{j}+(1-\zeta_{\alpha}^{j})\ln(1-\zeta_{\alpha}^{j})].\label{eq:EE12}
\end{equation}
Here $\{\zeta_{\alpha}^{j}|j=1,...,L\}$ represent the eigenvalues
of the single-particle correlation matrix
\begin{equation}
	C_{\alpha}^{\top}=\sum_{j=1}^{L}\zeta_{\alpha}^{j}|\phi_{\alpha}^{j}\rangle\langle\phi_{\alpha}^{j}|.\label{eq:CM}
\end{equation}
Its matrix elements are given by $(C_{\alpha})_{mn}=\langle\hat{\boldsymbol{{\bf c}}}_{m}^{\dagger}\hat{\boldsymbol{{\bf c}}}_{n}\rangle$
($m,n\in{\rm A}$), where the average is taken over the half-filled
initial state \cite{FloESEE}. $\{|\phi_{\alpha}^{j}\rangle\}$ are
the eigenvectors of $C_{\alpha}$. We will also refer to the set $\{\zeta_{\alpha}^{j}\}$
as the ES of $\hat{\rho}_{{\rm A}}^{\alpha}$ in the time frame $\alpha$.
Moreover, the topological properties of $\hat{H}_{{\rm A}}^{\alpha}$
can be characterized by its open-bulk winding number $W_{\alpha}$,
which is defined as \cite{FloESEE}
\begin{equation}
	W_{\alpha}=-\frac{1}{L'}{\rm Tr}'({\cal S}_{{\rm A}}Q_{\alpha}[Q_{\alpha},N_{{\rm A}}]).\label{eq:W12}
\end{equation}
Here the definitions of ${\cal S}_{{\rm A}}$, $N_{{\rm A}}$, $L'$
and ${\rm Tr}'$ are the same as those introduced in Eq.~(\ref{eq:W}).
The projector $Q_{\alpha}$ in the time frame $\alpha$ ($=1,2$)
is defined as
\begin{equation}
	Q_{\alpha}=\sum_{j=1}^{L}[\Theta(\zeta_{\alpha}^{j}-1/2)-\Theta(1/2-\zeta_{\alpha}^{j})]|\phi_{\alpha}^{j}\rangle\langle\phi_{\alpha}^{j}|.\label{eq:Qa}
\end{equation}
The bulk-edge correspondence of chiral symmetric 1D Floquet systems,
in either insulating or superconducting phases, can then be established
with the help of the ES and the $(W_{1},W_{2})$ from the quantum
entanglement viewpoint \cite{FloESEE}.

In the next section, we will see that even though the PKKC1 and PKKC2
share the same set of symmetries with the static KC, the
driving fields allow Majorana edge modes, topological phases and phase
transitions that are significantly different from and much richer
than those expected in the static KC to appear in these
Floquet models. The topological properties, entanglement nature and
bulk-edge correspondence of these intriguing nonequilibrium phases
will then be revealed.

\begin{table}
	\begin{centering}
		\begin{tabular}{|c|c|c|}
			\hline 
			Quantity & PKKC1 {[}Eq.~(\ref{eq:H1}){]} & PKKC2 {[}Eq.~(\ref{eq:H2}){]}\tabularnewline
			\hline 
			\hline
			Winding number & \multicolumn{2}{c|}{$w_{\alpha}=\int_{-\pi}^{\pi}\frac{dk}{4\pi}{\rm Tr}[{\cal S}{\cal Q}_{\alpha}(k)i\partial_{k}{\cal Q}_{\alpha}(k)]$}\tabularnewline
			\hline 
			Topological invariants & \multicolumn{2}{c|}{$(w_{0},w_{\pi})=\left(\frac{w_{1}+w_{2}}{2},\frac{w_{1}-w_{2}}{2}\right)$}\tabularnewline
			\hline 
			Entanglement spectrum & \multicolumn{2}{c|}{$\xi_{\alpha}^{j}=\ln(1/\zeta_{\alpha}^{j}-1)$}\tabularnewline
			\hline 
			Entanglement entropy & \multicolumn{2}{c|}{$S_{\alpha}=-\sum_{j}[\zeta_{\alpha}^{j}\ln\zeta_{\alpha}^{j}+(1-\zeta_{\alpha}^{j})\ln(1-\zeta_{\alpha}^{j})]$}\tabularnewline
			\hline 
			Entanglement winding number & \multicolumn{2}{c|}{$W_{\alpha}=-\frac{1}{L'}{\rm Tr}'({\cal S}_{{\rm A}}Q_{\alpha}[Q_{\alpha},N_{{\rm A}}])$}\tabularnewline
			\hline 
		\end{tabular}
		\par\end{centering}
	\caption{Various quantities and their definitions for the PKKC1
			and PKKC2. $\alpha=1,2$ are the indices of symmetric time frames.
			The ${\cal Q}_{\alpha}(k)$ is defined in Eq.~(\ref{eq:Q12}). The set $\{\zeta_{\alpha}^{j}\}$
			contains all the eigenvalues of the correlation matrix $C_{\alpha}$
			in Eq.~(\ref{eq:CM}). The meanings of $L'$, ${\rm Tr}'$, ${\cal S}_{A}$,
			$N_{A}$ and $Q_{\alpha}$ are discussed below Eq.~(\ref{eq:W12}). \label{tab:2}}
\end{table}

\section{Results}\label{sec:Res}
We now investigate the topological and entanglement features
of the two variants of periodically kicked KC introduced
in Eqs.~(\ref{eq:H1}) and (\ref{eq:H2}). In each case, we first
identify the topological phase diagram of the system under the PBC.
Next, we consider the Floquet spectrum of the system under the OBC
and build the correspondence between the bulk topological invariants
and the numbers of Floquet Majorana edge modes at zero and $\pi$
quasienergies. Finally, we extract the signatures of topological phases,
phase transitions and bulk-edge correspondence of the system from
its ES and EE under the PBC. We will see that the physical properties
of both models are drastically modified by driving fields.

\subsection{PKKC1: kicked pairing amplitude}\label{subsec:M1}
We start with the Floquet KC described by the Hamiltonian
$\hat{H}_{{\rm I}}(t)$ {[}Eq.~(\ref{eq:H1}){]} and the Floquet operator
$\hat{U}_{{\rm I}}$ {[}Eq.~(\ref{eq:U1}){]}. The bulk quasienergy
spectrum is obtained by solving the eigenvalue equation of $U_{{\rm I}}(k)$
in Eq.~(\ref{eq:U1k}), i.e., $U_{{\rm I}}(k)|\psi\rangle=e^{-iE}|\psi\rangle$,
yielding 
\begin{equation}
	\pm E(k)=\pm\arccos[\cos(\mu+J\cos k)\cos(\Delta\sin k)].\label{eq:E1k}
\end{equation}
We find that there are two Floquet bands, whose quasienergies are
symmetric with respect to $E=0$. Moreover, as $E(k)$ is defined
modulus $2\pi$, these bands are also symmetric with respect to $E=\pm\pi$.
The quasienergy spectrum could then possess gaps around $E=0$ and
$E=\pi$. When one or both of these quasienergy gaps close, we may
encounter a phase transition in the system. Requiring that $E(k)=0$
($\pi$), we find $\cos(\mu+J\cos k)\cos(\Delta\sin k)=1$ ($-1$),
yielding the unified phase boundary equation
\begin{equation}
	\frac{p^{2}\pi^{2}}{\Delta^{2}}+\frac{(q\pi-\mu)^{2}}{J^{2}}=1,\label{eq:PB1}
\end{equation}
with $p,q\in\mathbb{Z}$ under the condition that $|\Delta|\geq|p\pi|$
and $|\mu\pm J|\geq|q\pi|$. That is, the Floquet spectrum of our
PKKC1 will become gapless at $E=0$ or $\pi$ once its parameters
satisfy Eq.~(\ref{eq:PB1}). It is clear that the Eq.~(\ref{eq:PB1})
is rather different from and more complicated than the topological
phase boundary $J=\pm\mu$ of the static KC. The allowed
numbers of topological transitions could be much larger in the parameter
space $(J,\mu,\Delta)$, and the kicked pairing amplitude $\Delta$
may also play an important role in determining the topological phases.

\begin{figure}
	\begin{centering}
		\includegraphics[scale=0.75]{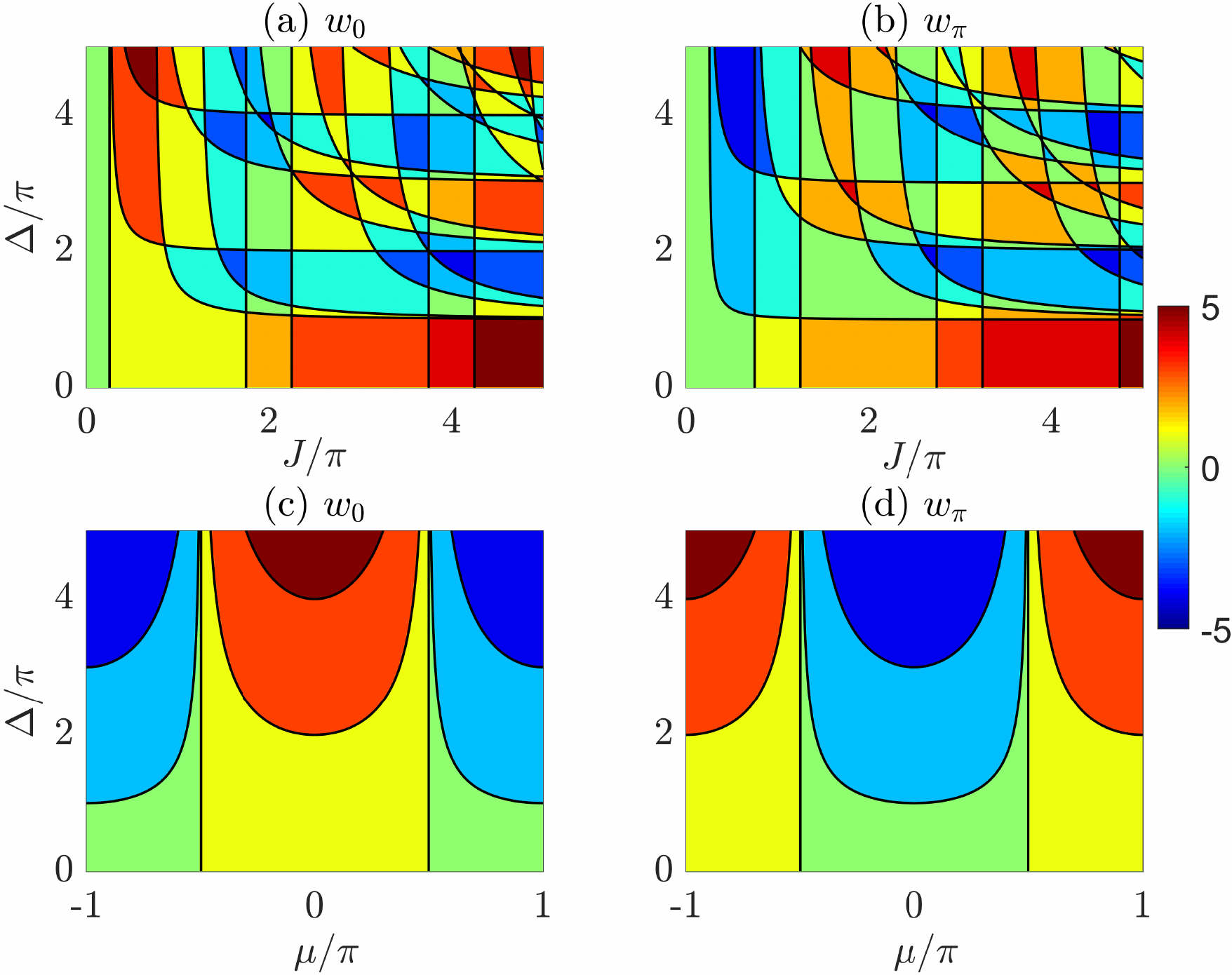}
		\par\end{centering}
	\caption{Topological phase diagrams of the PKKC1. (a) and (b) show the winding
		numbers $w_{0}$ and $w_{\pi}$ versus $(J,\Delta)$ for $\mu=\pi/4$.
		(c) and (d) show the winding numbers $w_{0}$ and $w_{\pi}$ versus
		$(\mu,\Delta)$ for $J=\pi/2$. In each panel, every region with a
		uniform color corresponds to a phase in which the winding number $w_{0}$
		or $w_{\pi}$ takes a quantized value, as can be read out from the
		common color bar of the four panels. \label{fig:PD1}}
\end{figure}

In Fig.~\ref{fig:PD1}, we present the topological phase diagrams
of the PKKC1. They are obtained by evaluating the winding numbers
$(w_{0},w_{\pi})$ with the help of Eqs.~(\ref{eq:Uk})--(\ref{eq:w0p}),
where we have $h_{a}(k)=\Delta\sin k\sigma_{y}$ and $h_{b}(k)=(\mu+J\cos k)\sigma_{z}$
according to the Eq.~(\ref{eq:U1k}). The boundary lines between different
topological phases are consistent with the prediction of Eq.~(\ref{eq:PB1}).
Compared with the Fig.~\ref{fig:KC}(a), we find that the PKKC1
indeed owns rich patterns of topological phases and phase transitions
in a broad range of the hopping amplitude $J$, pairing strength $\Delta$
and chemical potential $\mu$. A couple of points deserve to be mentioned.
First, in certain parameter regions {[}e.g., $\Delta\in(0,\pi/2)$
in Figs.~\ref{fig:PD1}(a) and \ref{fig:PD1}(b) or $\mu=0$ in Figs.~\ref{fig:PD1}(c) and \ref{fig:PD1}(d){]}, both the $w_{0}$
and $w_{\pi}$ can increase monotonically with the increase of $J$
or $\Delta$. This means that we can in principle generate arbitrarily
many phase transitions and obtain Floquet superconducting phases with
arbitrarily large topological winding numbers by varying a single
system parameter in the PKKC1, which is obviously not achievable in
its static counterpart. Second, the change of pairing amplitude $\Delta$
could also guide the system to roam between different topological
phases. This indicates that the p-wave pairing could play a more active
role in controlling the topological properties of Floquet KC
in comparison with what it did in the static case. Third, in the topological
trivial regime ($|J|<|\mu|$) of the static KC, our kicked
model still holds many topological nontrivial phases. Their presence
demonstrates one key advantage of Floquet engineering, i.e., to transform
a trivial static phase to multiple nontrivial Floquet topological
phases.

\begin{figure}
	\begin{centering}
		\includegraphics[scale=0.75]{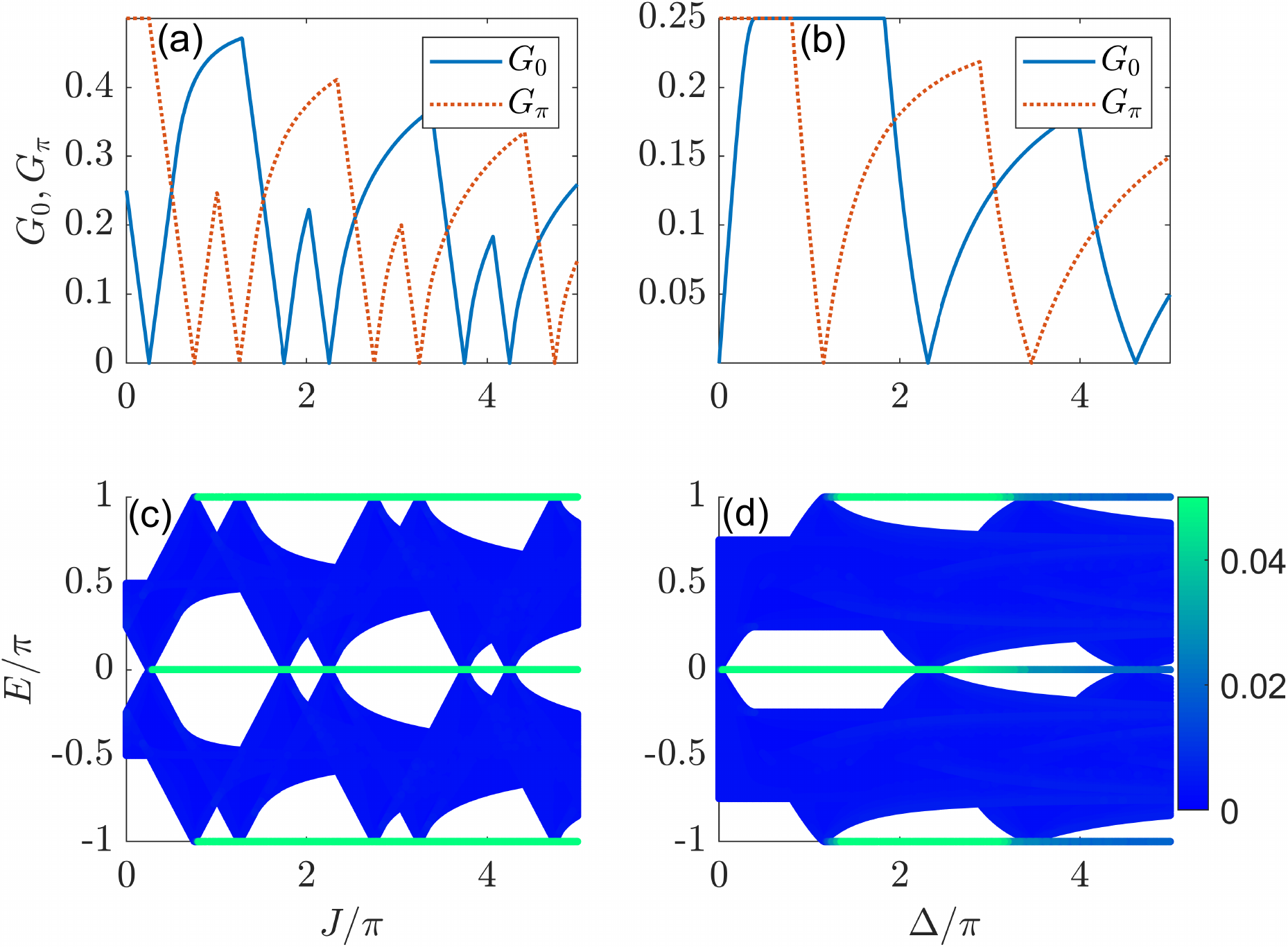}
		\par\end{centering}
	\caption{Gap functions under the PBC {[}(a) and (b){]} and quasienergy spectra
		under the OBC {[}(c) and (d){]} of the PKKC1 versus $J$ and $\Delta$.
		Other system parameters are $(\Delta,\mu)=(\pi/2,\pi/4)$ for (a),
		(c) and $(J,\mu)=(\pi/2,\pi/4)$ for (b), (d). (c) and (d) share the
		same color bar, and the color of each data point represents the IPR
		of the corresponding Floquet eigenstate \cite{Note1}. \label{fig:EOBC1}}
\end{figure}

In Figs.~\ref{fig:EOBC1}(a) and \ref{fig:EOBC1}(b), we show the
gap functions $G_{0}$ and $G_{\pi}$ of our PKKC1 under the PBC versus
the hopping and pairing amplitudes $J$ and $\Delta$, respectively.
$G_{0}$ and $G_{\pi}$ are defined as
\begin{equation}
	G_{0}=\frac{1}{\pi}\min_{k\in{\rm BZ}}|E(k)|,\qquad G_{\pi}=\frac{1}{\pi}\min_{k\in{\rm BZ}}|E(k)-\pi|,\label{eq:G0P}
\end{equation}
where the $E(k)$ is given by Eq.~(\ref{eq:E1k}). It is clear that
we will have $G_{0}=0$ ($G_{\pi}=0$) if and only if the quasienergy
gap closes at $E=0$ ($E=\pi$). The locations of these gapless points
in the parameter space are coincident with the predictions of 
Eq.~(\ref{eq:PB1}) and the topological phase diagrams in Fig.~\ref{fig:PD1}.
To check whether these transitions accompany the changes of Floquet
Majorana edge modes, we can investigate the quasienergy spectrum and
states of the system under the OBC. According to the Eqs.~(\ref{eq:BdG3}),
(\ref{eq:BdG4}) and (\ref{eq:UBdG}), the matrices $\mathsf{H}_{a}$
and $\mathsf{H}_{b}$ for our PKKC1 in the basis $(u_{1},v_{1},...,u_{n},v_{n},...,u_{L},v_{L})^{{\rm \top}}$
are given by $\mathsf{H}_{a}=\frac{\Delta}{2i}\sum_{n}(|n\rangle\langle n+1|-{\rm H.c.})\otimes\sigma_{y}$
and $\mathsf{H}_{b}=\frac{1}{2}\sum_{n}(\mu|n\rangle\langle n|+J|n\rangle\langle n+1|+{\rm H.c.})\otimes\sigma_{z}$.
Solving the eigenvalue equation for $\mathsf{U}$ in the Eq.~(\ref{eq:UBdG}),
i.e., $\mathsf{U}|\psi\rangle=e^{-iE}|\psi\rangle$ under the OBC,
we can find the Floquet spectrum of the system. Two typical examples
of the spectrum are shown in Figs.~\ref{fig:EOBC1}(c) and \ref{fig:EOBC1}(d).
We observe that with the increase of $J$ or $\Delta$, the system
could indeed undergo a series of gap-closing transitions at $E=0$
and $E=\pi$. Moreover, many Floquet Majorana edge modes emerge at
these quasienergies following the transitions. In each gapped phase,
a simple counting reveals that the numbers $(n_{0},n_{\pi})$ of Floquet
Majorana zero and $\pi$ edge modes are related to the winding numbers
$(w_{0},w_{\pi})$ in Fig.~\ref{fig:PD1} through the relations $n_{0}=2|w_{0}|$
and $n_{\pi}=2|w_{\pi}|$, which describe the bulk-edge correspondence
of our PKKC1. Notably, we find that both the numbers of zero and $\pi$
Floquet Majorana edge modes could increase monotonically with the
increase of $J$ or $\Delta$. Therefore, we can in principle obtain
arbitrarily many Majorana zero and $\pi$ modes in the limit $L\rightarrow\infty$
by tuning a single parameter of the system, which is not achievable
for the static KC. These abundant Majorana modes may provide
more room for the realization of Floquet topological quantum computation
\cite{FloQC4}.

\begin{figure}
	\begin{centering}
		\includegraphics[scale=0.75]{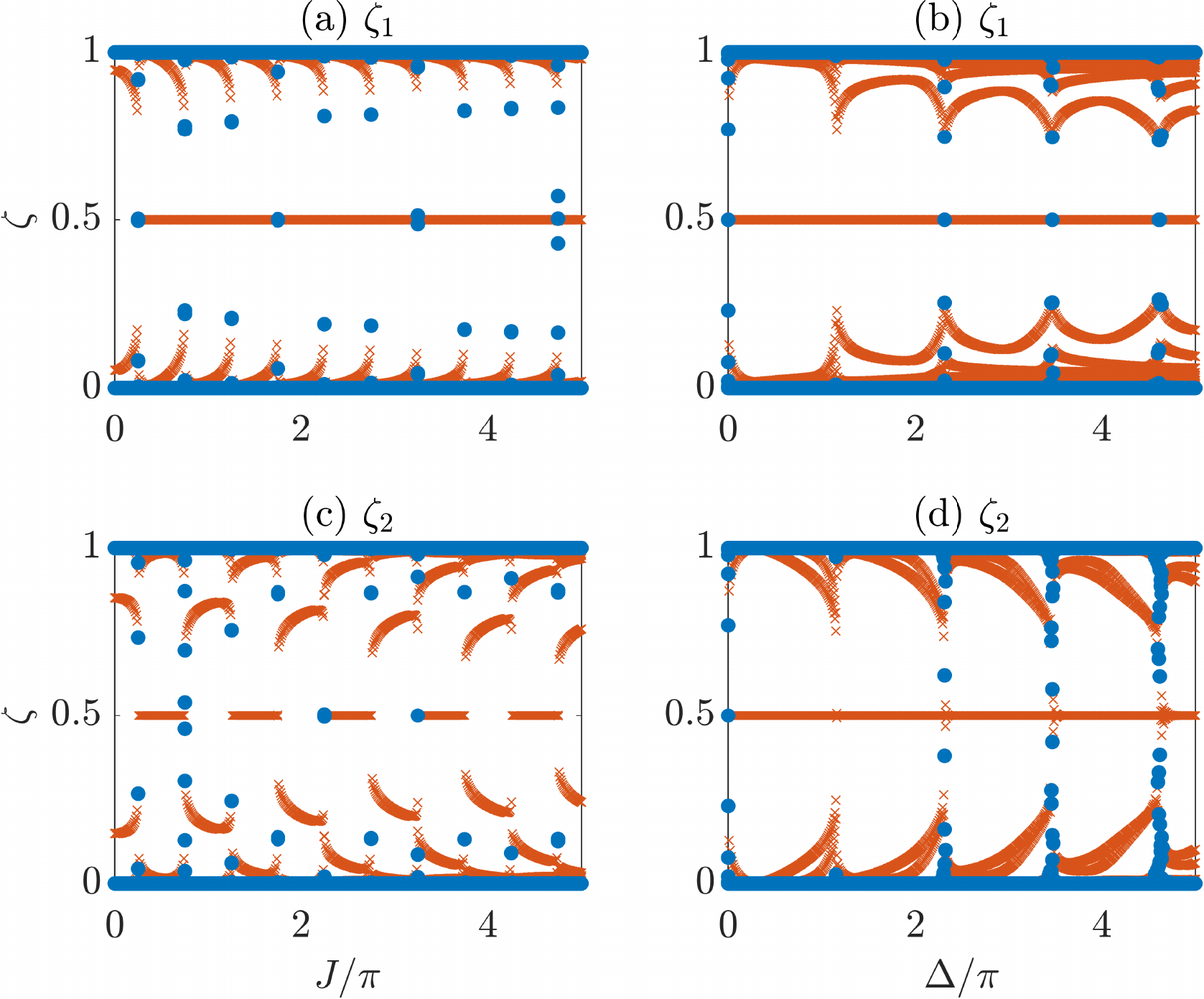}
		\par\end{centering}
	\caption{ES of the PKKC1 versus $J$ {[}(a), (c){]} and $\Delta$ {[}(b), (d){]}
		in different time frames. Other system parameters are $(\Delta,\mu)=(\pi/2,\pi/4)$
		for (a), (c) and $(J,\mu)=(\pi/2,\pi/4)$ for (b), (d). $\zeta_{1}$
		and $\zeta_{2}$ denote the ES {[}Eq.~(\ref{eq:ES12}){]} obtained
		in the first and second symmetric time frames, respectively. The crosses
		highlight localized entanglement eigenmodes in the subsystem A. \label{fig:ES1}}
\end{figure}

\begin{figure}
	\begin{centering}
		\includegraphics[scale=0.75]{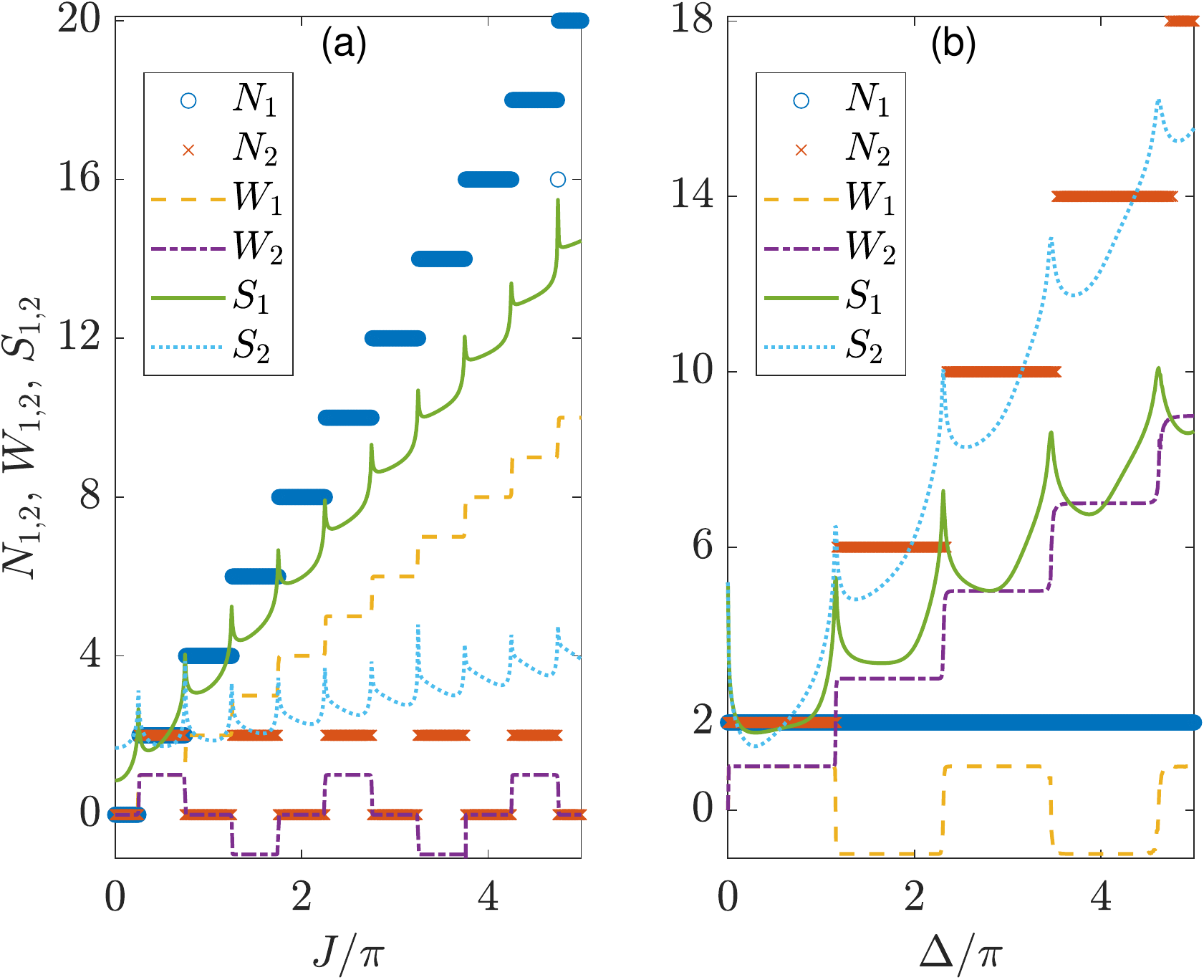}
		\par\end{centering}
	\caption{The numbers $N_{1,2}$ of $\zeta=1/2$ eigenmodes in the ES, the winding
		numbers $W_{1,2}$ of entanglement Hamiltonians $\hat{H}_{{\rm A}}^{1,2}$
		and the EE $S_{1,2}$ of the PKKC1 versus $J$ and $\Delta$ in two
		symmetric time frames. Other system parameters are $(\Delta,\mu)=(\pi/2,\pi/4)$
		for (a) and $(J,\mu)=(\pi/2,\pi/4)$ for (b). The calculations are
		done for the system at half-filling and under the PBC. \label{fig:EE1}}
\end{figure}

We now investigate the topology and bulk-edge correspondence of the
PKKC1 from the entanglement perspective. Referring to Eqs.~(\ref{eq:HBdG1})
and (\ref{eq:UESEE}), we can identify the $\hat{{\cal H}}_{a}$ and
$\hat{{\cal H}}_{b}$ of our PKKC1 as 
\begin{equation}
	\hat{{\cal H}}_{a}=\frac{\Delta}{2i}\sum_{n}(\hat{\boldsymbol{{\bf c}}}_{n}^{\dagger}\sigma_{y}\hat{\boldsymbol{{\bf c}}}_{n+1}-{\rm H.c.}),\label{eq:Ha1}
\end{equation}
\begin{equation}
	\hat{{\cal H}}_{b}=\frac{1}{2}\sum_{n}(\mu\hat{\boldsymbol{{\bf c}}}_{n}^{\dagger}\sigma_{z}\hat{\boldsymbol{{\bf c}}}_{n}+J\hat{\boldsymbol{{\bf c}}}_{n}^{\dagger}\sigma_{z}\hat{\boldsymbol{{\bf c}}}_{n+1}+{\rm H.c.}).\label{eq:Hb1}
\end{equation}
Plugging them into the expressions of $\hat{{\cal U}}_{1}$ and $\hat{{\cal U}}_{2}$
in Eqs.~(\ref{eq:UI}) and (\ref{eq:UII}), we obtain the Floquet
operators of the system described by $\hat{{\cal H}}_{{\rm I}}(t)$
{[}Eq.~(\ref{eq:HBdG1}){]} in two symmetric time frames. Considering
a spatially bisected, half-filled lattice under the PBC and following
the procedure discussed in the last section, we can obtain the bipartite
ES $\zeta_{\alpha}$ and EE $S_{\alpha}$ of $\hat{{\cal U}}_{\alpha}$
in the symmetric time frame $\alpha$ for $\alpha=1,2$. Their combination
could generate a complete topological characterization for the PKKC1.
In Fig.~\ref{fig:ES1}, we show the changes of $(\zeta_{1},\zeta_{2})$
with respect to $J$ and $\Delta$ for two typical situations. We
observe that in both cases, the configuration of either $\zeta_{1}$
or $\zeta_{2}$ changes discontinuously around $\zeta=1/2$ whenever
the full system described by $\hat{U}_{{\rm I}}$ {[}Eq.~(\ref{eq:U1}){]}
undergoes a topological phase transition as observed in Fig.~\ref{fig:PD1}.
A direct counting of the $\zeta=1/2$ entanglement eigenmodes in the
two symmetric time frames results in the $N_{1}$ and $N_{2}$ in
Fig.~\ref{fig:EE1}. We observe that within each topological phase
of our PKKC1, the sum and difference of $N_{1}$ and $N_{2}$ do not change their values,
while either or both of them get quantized jumps when the system goes
through a topological phase transition point. These $\zeta=1/2$ eigenmodes
are localized around the entanglement cuts between the subsystems
A and B. Each of them contributes a $\Delta S=\ln2$ to the EE 
{[}Eq.~(\ref{eq:EE12}){]}. We can thus view them as edge states of the Floquet
entanglement Hamiltonian $\hat{H}_{{\rm A}}^{\alpha}$ in the time
frame $\alpha$, similar to the case of the static KC. The
numbers of these entanglement edge modes are further expected to be
related to the Floquet Majorana zero and $\pi$ edge modes in the
parent system.

The EE {[}Eq.~(\ref{eq:EE12}){]} and the winding numbers {[}Eq.~(\ref{eq:W12}){]}
of Floquet entanglement Hamiltonians are also shown in the Fig.~\ref{fig:EE1}.
In both time frames $\alpha=1,2$, we find non-analytic cusps in the
EE whenever the system undergoes a topological phase transition as
observed in Figs.~\ref{fig:PD1} and \ref{fig:EOBC1}. The winding
numbers $(W_{1},W_{2})$ are obtained by inserting the Eqs.~(\ref{eq:Ha1})
and (\ref{eq:Hb1}) into Eq.~(\ref{eq:UESEE}) and following the steps
from Eqs.~(\ref{eq:UI}) to (\ref{eq:Qa}). The results show that
the winding numbers $W_{1}$ and $W_{2}$ also get integer jumps at
the transition points, confirming their topological nature from an
entanglement viewpoint. Moreover, $W_{1}$ and $W_{2}$ remain quantized
within each topological phase, and they are related to the numbers
of $\zeta=1/2$ entanglement eigenmodes through the relations
\begin{equation}
	N_{1}=2|W_{1}|,\qquad N_{2}=2|W_{2}|.\label{eq:BBC1}
\end{equation}
We refer to the Eq.~(\ref{eq:BBC1}) as the bulk-edge correspondence
in the ES of our PKKC1. A further comparison between the bulk winding
numbers $(w_{1},w_{2})$ in the Eq.~(\ref{eq:w12}) and the entanglement
winding numbers $(W_{1},W_{2})$ suggest that $(w_{1},w_{2})=(W_{1},W_{2})$.
This observation provides important insights for the physical meaning
of $w_{\alpha}$ in Eq.~(\ref{eq:w12}), i.e., it counts the number
of $\zeta=1/2$ eigenmodes of the Floquet entanglement Hamiltonian
in the symmetric time frame $\alpha$. We expect this to
hold not only for Floquet topological superconductors but also for
Floquet topological insulators in one-dimension with chiral symmetry.
Finally, we can write down the correspondence between the entanglement
winding numbers $(W_{1},W_{2})$ and the numbers of Floquet Majorana
edge modes $(n_{0},n_{\pi})$ in the quasienergy spectrum, i.e.,
\begin{equation}
	n_{0}=|W_{1}+W_{2}|,\qquad n_{\pi}=|W_{1}-W_{2}|.\label{eq:EBBC1}
\end{equation}
Therefore, the ES and EE could also provide us with sufficient information
to identify topological phase transitions, characterizing different
topological phases and describing the bulk-edge correspondence of 1D
Floquet topological superconductors. We will further confirm this
view by investigating the Floquet KC under a different kicking
protocol in the following subsection.

\subsection{PKKC2: kicked hopping amplitude}\label{subsec:M2}

We now consider the kicked KC described by the Hamiltonian
$\hat{H}_{{\rm II}}(t)$ {[}Eq.~(\ref{eq:H2}){]} and Floquet
operator $\hat{U}_{{\rm II}}$ {[}Eq.~(\ref{eq:U2}){]}. 
The calculation details can be worked out in parallel with the PKKC1, and they are explicitly given in Appendix \ref{sec:APP3}.
The equations that describing all the phase boundaries of the PKKC2 can be found analytically. They are clearly different from those depicted
by either Eq.~(\ref{eq:PB1}) for the PKKC1 or the phase boundaries
of the static KC. Meanwhile, the pairing amplitude $\Delta$
could again play important roles in controlling phase transitions
in the system, rather than serving passively as a unit of energy in
the static KC.

\begin{figure}
	\begin{centering}
		\includegraphics[scale=0.75]{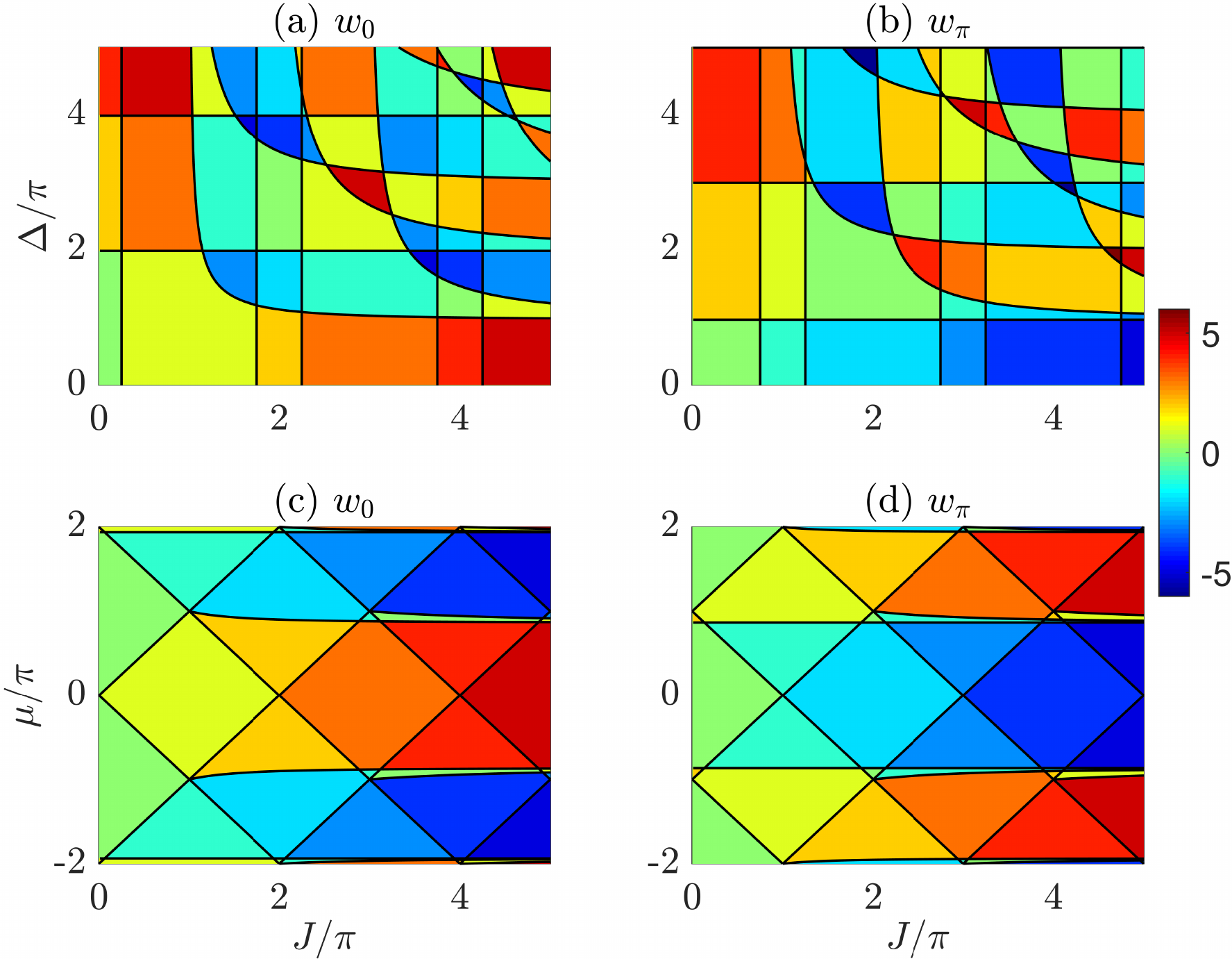}
		\par\end{centering}
	\caption{Topological phase diagrams of the PKKC2. (a) and (b) present the winding
		numbers $w_{0}$ and $w_{\pi}$ versus $(J,\Delta)$ for $\mu=\pi/4$.
		(c) and (d) present the winding numbers $w_{0}$ and $w_{\pi}$ versus
		$(J,\mu)$ for $\Delta=\pi/2$. In each panel, every region with the
		same color corresponds to a phase in which $w_{0}$ or $w_{\pi}$
		takes the same integer value, as can be figured out from the shared
		color bar of the four panels. \label{fig:PD2}}
\end{figure}

In Fig.~\ref{fig:PD2}, we give examples of topological
phase diagrams of the PKKC2. We observe again rich Floquet
topological superconducting phases and transitions
induced by the driving field in different parameter regions. The boundaries
(in black lines) across which $w_{0}$ and $w_{\pi}$ get quantized
changes are correctly captured by the Eqs.~(\ref{eq:PB21}) and (\ref{eq:PB22}).
Moreover, the values of $w_{0}$ and $w_{\pi}$ can both grow monotonically
with the increase of $J$ or $\Delta$, leading to Floquet topological
phases with arbitrarily huge winding numbers under large hopping or
pairing amplitudes, which cannot be reached in the static KC
in similar parameter regions. The reason behind the emergence of
these large winding number phases in both the PKKC1 and PKKC2 is the
effective long-range coupling generated by the piecewise Floquet evolution
in Eqs.~(\ref{eq:U1}) and (\ref{eq:U2}). We can infer this directly
from the bulk dispersion relations in Eqs.~(\ref{eq:E1k}) and (\ref{eq:E2k}),
where long-range hopping and pairing terms in real-space are expected
upon inverse Fourier transformations. In the meantime, we also observe
many topological nontrivial phases and transitions in the regime $|J|<|\mu|$,
where the static KC is topologically trivial. Therefore,
time-periodic kickings applied to either the hopping
or pairing amplitudes of KC could significantly alter
its topological properties.

\begin{figure}
	\begin{centering}
		\includegraphics[scale=0.75]{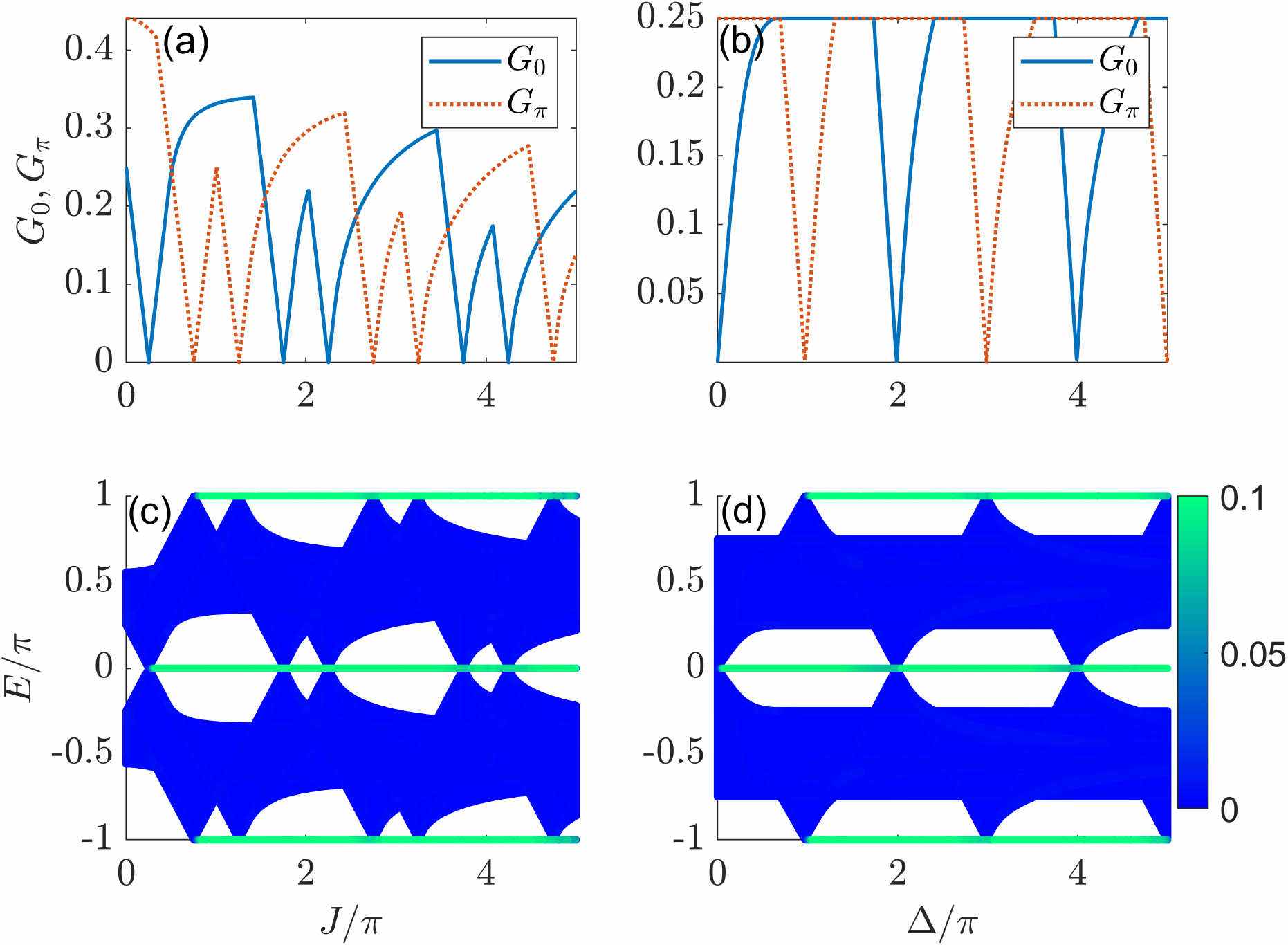}
		\par\end{centering}
	\caption{Gap functions under the PBC {[}(a) and (b){]} and quasienergy spectra
		under the OBC {[}(c) and (d){]} of the PKKC2 versus $J$ and $\Delta$.
		Other system parameters are $(\Delta,\mu)=(\pi/2,\pi/4)$ for (a),
		(c) and $(J,\mu)=(\pi/2,\pi/4)$ for (b), (d). (c) and (d) have the
		same color bar, and the color of each data point is given by the IPR of
		the corresponding Floquet state \cite{Note1}. \label{fig:EOBC2}}
\end{figure}

In Fig.~\ref{fig:EOBC2}, we show the quasienergy gap functions 
{[}Eq.~(\ref{eq:G0P}){]} under the PBC and Floquet spectra under the OBC
for two typical cases of PKKC2. 
We observe that in the parameter space of $J$ and $\Delta$, the
locations where the gap function $G_{0}=0$ ($G_{\pi}=0$) under the
PBC are precisely consistent with the positions where the quasienergy
gap closes at $E=0$ ($E=\pi$) under the OBC. Moreover, the two types
of degenerate Majorana modes appearing at $E=0$ and $E=\pi$ in the
gapped regime are both localized around the two edges of the lattice.
The numbers of these Majorana edge modes $(n_{0},n_{\pi})$ and their
changes versus $(J,\Delta)$ are found to be precisely captured by
the winding numbers $(w_{0},w_{\pi})$ in the bulk topological phase
diagram (Fig.~\ref{fig:PD2}) through the bulk-edge correspondence
\begin{equation}
	n_{0}=2|w_{0}|,\qquad n_{\pi}=2|w_{\pi}|.\label{eq:BEC}
\end{equation}
It is expected that these relations for the PKKC1 and PKKC2 are identical,
as both models are in one-dimension with the same set of protecting
symmetries and they are thus characterized by the same type of topological
winding numbers. Meanwhile, we find a series of topological phase
transitions with the increase of $J$ and $\Delta$, after which more
and more Floquet Majorana zero and $\pi$ edge modes tend to appear
in the spectral gaps. Therefore, similar to the case of PKKC1, we
can obtain Floquet superconducting phases with arbitrarily many
Majorana edge modes in principle in the PKKC2.

\begin{figure}
	\begin{centering}
		\includegraphics[scale=0.75]{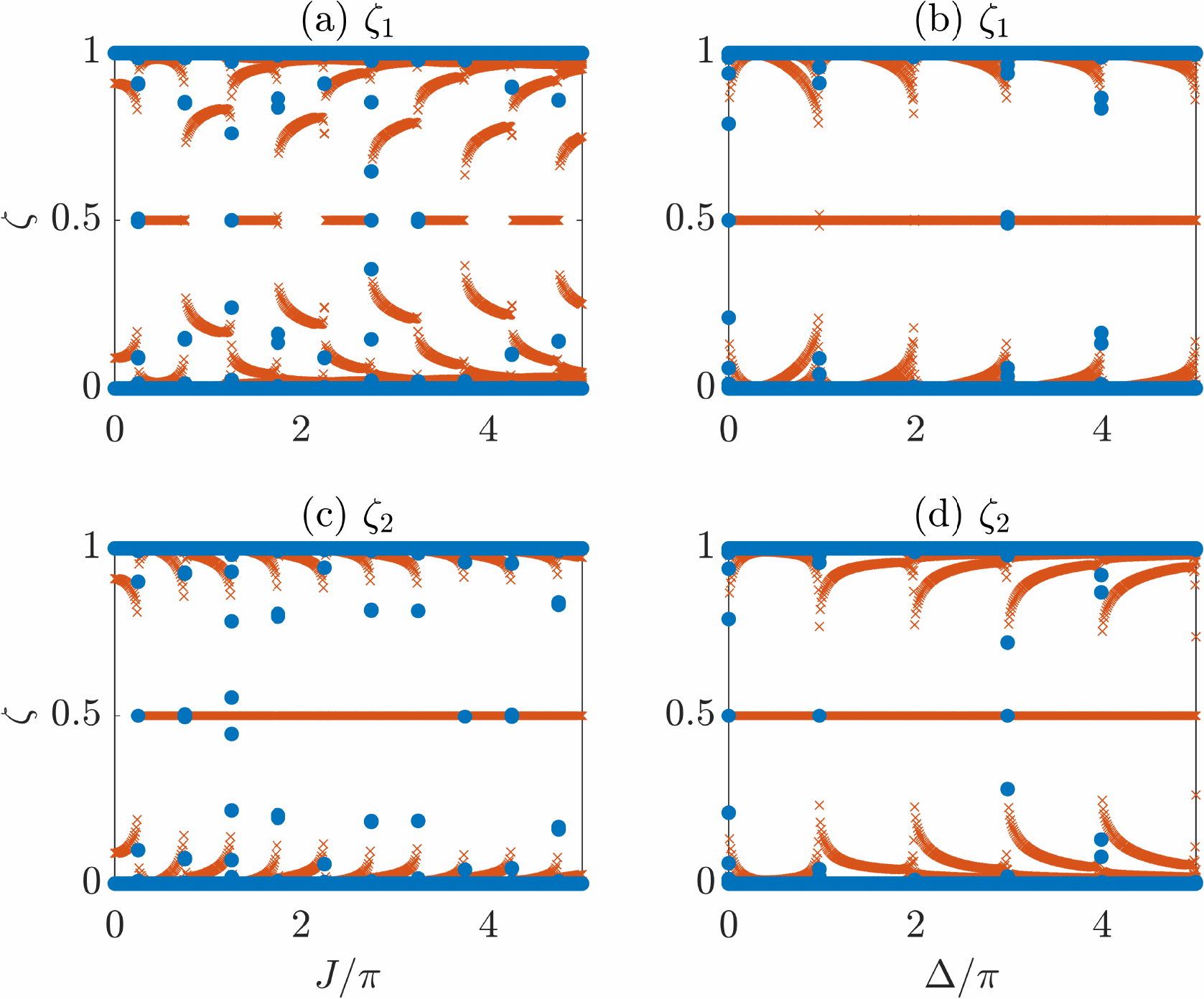}
		\par\end{centering}
	\caption{ES of the PKKC2 versus $J$ {[}(a), (c){]} and $\Delta$ {[}(b), (d){]}
		under the PBC and at half-filling. Other system parameters are
		$(\Delta,\mu)=(\pi/2,\pi/4)$ for (a), (c) and $(J,\mu)=(\pi/2,\pi/4)$
		for (b), (d). $\zeta_{1}$ and $\zeta_{2}$ represent the ES evaluated
		in the first and second time frames, respectively. The crosses
		highlight localized entanglement eigenmodes in the subsystem A. \label{fig:ES2}}
\end{figure}

\begin{figure}
	\begin{centering}
		\includegraphics[scale=0.75]{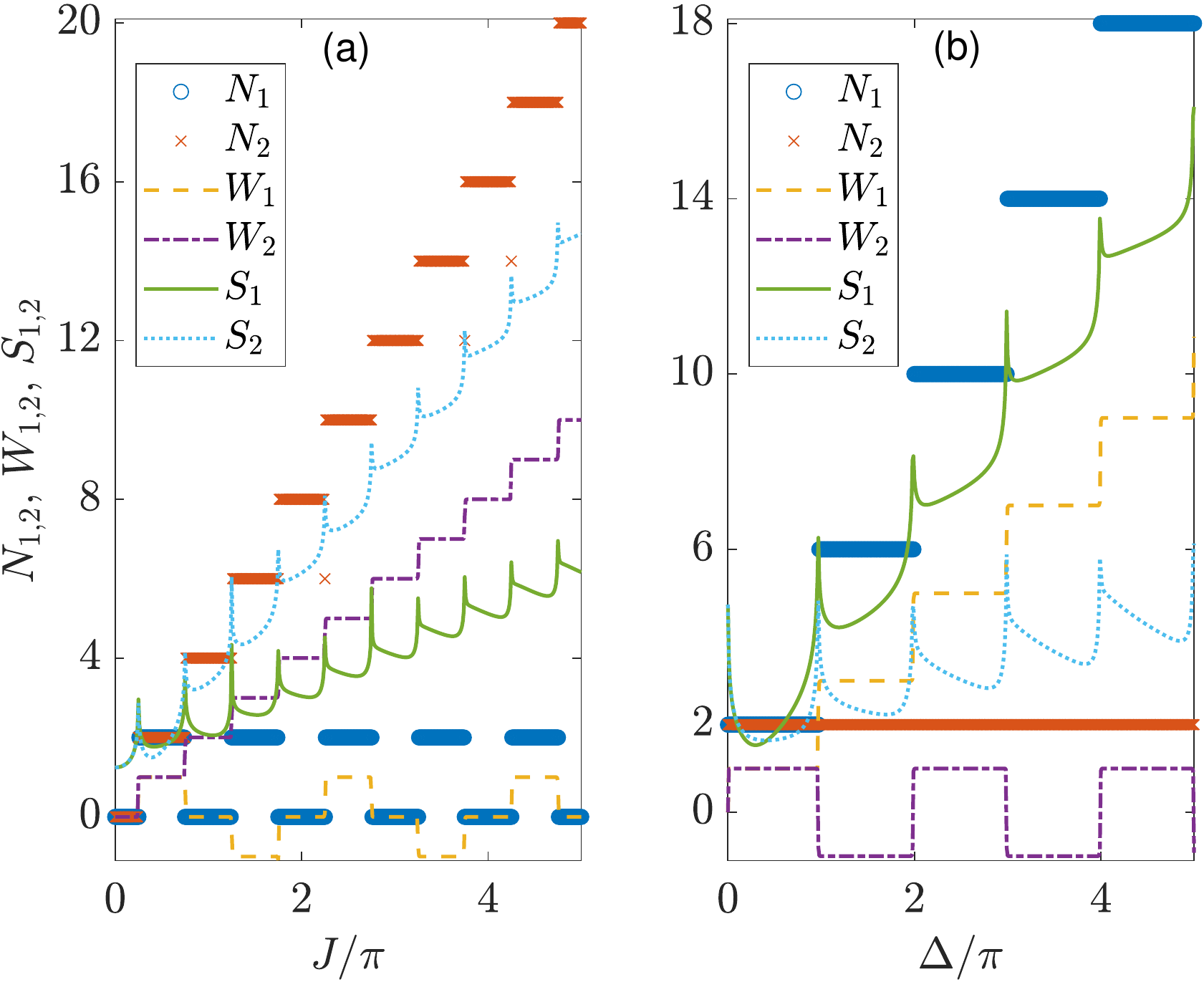}
		\par\end{centering}
	\caption{The numbers $N_{1,2}$ of $\zeta=1/2$ eigenmodes in the ES, the winding
		numbers $W_{1,2}$ of entanglement Hamiltonians and the EE $S_{1,2}$
		versus $J$ and $\Delta$ of the PKKC2 in two symmetric time frames. Other system parameters
		are set to $(\Delta,\mu)=(\pi/2,\pi/4)$ for (a) and $(J,\mu)=(\pi/2,\pi/4)$
		for (b). The calculations are done for the system at half-filling
		and under the PBC. \label{fig:EE2}}
\end{figure}

In Fig.~\ref{fig:ES2}, we present the ES of the PKKC2 in two symmetric
time frames within the same parameter ranges of Fig.~\ref{fig:EOBC2}.
Similar to the case of PKKC1, we observe eigenmodes
in the ES with $\zeta=1/2$ for the system in both time frames. These
eigenmodes are localized around the entanglement cuts between the
subsystems A and B, with each of them contributing a $\Delta S=\ln2$
to the EE in Fig.~\ref{fig:EE2}. Furthermore, the numbers of these
entanglement edge modes $(N_{1},N_{2})$ are found to undergoing quantized
changes whenever a topological phase transition happens with the change
of $J$ or $\Delta$, as shown in Fig.~\ref{fig:EE2}. A simple counting
suggests the following relations between the numbers of Floquet Majorana
edge modes of $\hat{U}_{{\rm II}}$ and the numbers of its entanglement
edge modes with $\zeta=1/2$, i.e., 
\begin{equation}
	\max(N_{1},N_{2})=n_{0}+n_{\pi},\qquad\min(N_{1},N_{2})=|n_{0}-n_{\pi}|.\label{eq:N12n0p}
\end{equation}
We have systematically verified these relations for both the PKKC1
and PKKC2 in our numerical calculations (see also Appendix \ref{sec:App2}), which implies that they represent
a general set of equalities between the spectral and entanglement
topology of 1D, chiral symmetric Floquet topological superconductors.

Finally, we show in Fig.~\ref{fig:EE2} the number of $\zeta=1/2$
entanglement eigenmodes, the EE {[}Eq.~(\ref{eq:EE12}){]} and the
winding numbers {[}Eq.~(\ref{eq:W12}){]} of Floquet entanglement
Hamiltonians in two symmetric time frames $\alpha=1,2$ for the PKKC2.
At every transition point predicted by the Eqs.~(\ref{eq:PB21}) and
(\ref{eq:PB22}), we observe non-analytic cusps in both $S_{1}$ and
$S_{2}$. Therefore, we can extract the signatures of Floquet topological
phase transitions from the bipartite EE of our system. In combination
with the results for the PKKC1, we realize that the connection between
the non-analyticity of EE and the phase transitions in 1D, chiral
symmetric Floquet topological superconductors should be general and
independent of the exact forms of driving protocols. Furthermore, whenever the system described by Eq.~(\ref{eq:U2}) undergoes
a topological phase transition, we have an integer-quantized jump
in $W_{1}$ and/or $W_{2}$. In parameter regions where the quasienergy
spectrum of PKKC2 is gapped,
both $W_{1}$ and $W_{2}$ stay on quantized plateaus. Moreover, a
simple comparison implies that their values are related to the numbers
of $\zeta=1/2$ entanglement eigenmodes $(N_{1},N_{2})$ through the
Eq.~(\ref{eq:BBC1}). Therefore, Eq.~(\ref{eq:BBC1}) describes the
entanglement bulk-edge correspondence in both the PKKC1 and PKKC2,
which suggests that it holds more generally for 1D Floquet topological
superconductors in the BDI symmetry class. Referring to the winding
numbers of Eq.~(\ref{eq:U2}) that are reported in Fig.~\ref{fig:PD2},
we further notice that $W_{\alpha}=w_{\alpha}$ for $\alpha=1,2$.
Therefore, the winding numbers $w_{\alpha}$ as defined in Eq.~(\ref{eq:w12})
counts the number of $\zeta=1/2$ entanglement edge modes in the time
frame $\alpha$ for $\alpha=1,2$, instead of the zero or $\pi$
quasienergy edge modes in the corresponding time frame. The same relation
holds also for the PKKC1. To the best of our knowledge, this is the
first time that a concrete bulk-edge correspondence is identified
directly for the winding numbers $(w_{1},w_{2})$ of chiral symmetric
Floquet systems in symmetric time frames. It highlights the indispensable
role of quantum entanglement for our understanding of Floquet topological
matter. To accomplish the discussion, we notice that the relation
in Eq.~(\ref{eq:EBBC1}) between Floquet Majorana zero/$\pi$ edge
modes and winding numbers $(W_{1},W_{2})$ is also satisfied for
the PKKC2. This observation suggests that regardless of the periodic
driving protocols, the entanglement winding numbers $(W_{1},W_{2})$
could provide us with a complete topological characterization of 1D
Floquet superconducting phases in the BDI symmetry class.

\section{Discussion and conclusion}\label{sec:Sum}

In this work, by applying time-periodic kickings to either the hopping
or pairing amplitudes of the Kitaev chain, we found 1D Floquet superconducting
phases with large topological invariants and many Majorana edge modes.
Under the PBC, we analytically obtained the phase boundaries of the
system for each driving protocol, and numerically found the topological
winding numbers $(w_{0},w_{\pi})$ that characterizing each Floquet
superconducting phase of the bulk. Under the OBC, we found many Majorana
edge modes with zero and $\pi$ quasienergies in the Floquet spectrum.
The correspondence between their numbers and the values of $(w_{0},w_{\pi})$
in each gapped topological phase was also established. Patterns of
topological phases and transitions that are rather different from
and much richer than those in the static Kitaev chain were generated
under both kicking protocols thanks to the applied driving fields.
Finally, using the ES, EE and winding numbers of Floquet entanglement
Hamiltonians, we provided a complete characterization of the topological
phases and bulk-edge correspondence for 1D, chiral symmetric Floquet
superconducting systems from the entanglement perspective. Our work
thus unveiled the potential of Floquet engineering in producing Majorana
zero and $\pi$ edge modes as many as possible, which may find applications
in realizing Majorana time crystals and Floquet quantum computing
\cite{FloQC3,FloQC4,FloQC5}. Moreover, the entanglement-based framework
of characterizing Floquet topological superconductors could offer
us more insights about the physical meaning and implication of topological
invariants in Floquet systems, which may find extensions to other
symmetry classes and higher spatial dimensions. 

Theoretically, even though the topological and entanglement bulk-edge correspondences in Eqs.~(\ref{eq:BBC1}), (\ref{eq:EBBC1}), (\ref{eq:BEC}) and (\ref{eq:N12n0p}) are extracted from numerical calculations, they are expected to hold in any 1D Floquet topological insulators and superconductors in the symmetry classes AIII and BDI~\cite{FloCls3}, which are both characterized by the integers $(w_0,w_\pi)$ in Eq.~(\ref{eq:w0p}). With slight modifications, these relations should also be applicable to 1D Floquet models in the symmetry class CII, which are characterized by $2\mathbb{Z}\times2\mathbb{Z}$ topological invariants \cite{FloESEE}. We thus expect that our identifications about the entanglement winding numbers, the topological winding numbers and the numbers of edge modes are valid in generic 1D gapped Floquet topological phases protected by chiral symmetry, which go beyond the models considered in the present work.

Several aspects of the entanglement perspective, which offer information beyond the winding numbers of Floquet operators deserve to be emphasized. First, the EE shows a non-analytic cusp whenever the number of Majorana zero or $\pi$ edge modes changes. It could thus offer a sharp signal for the phase transitions in Floquet topological superconductors from the quantum information viewpoint, suggesting alternative routes of detecting these transitions. Second, our results revealed a ``bulk-edge correspondence'' [Eq.~(\ref{eq:BBC1})] between the number of localized eigenmodes around the entanglement cuts in the ES and the winding number of the Floquet entanglement Hamiltonian in each symmetric time frame, which could not be directly captured by the quasienergy spectra of Floquet operators. Third, our entanglement-based calculation endued a concrete physical meaning with the winding numbers $(w_1,w_2)$ [Eq.~(\ref{eq:w12})]. These winding numbers are previously employed only as intermediate quantities to define the invariants $(w_0,w_\pi)$ [Eq.~(\ref{eq:w0p})]. In the time frame $\alpha$, one could not directly predict the numbers of Floquet zero and $\pi$ edge modes under the OBC by just calculating $w_\alpha$ ($\alpha=1,2$). However, we unveiled that the momentum-space winding number $w_\alpha$ is actually equal to the real-space winding number $W_\alpha$ [Eq.~(\ref{eq:W12})] of the entanglement Hamiltonian, with the latter counting the number of entanglement edge modes  [Eq.~(\ref{eq:BBC1})]. Therefore, we established an ``entanglement bulk-edge correspondence'' for $w_\alpha$, i.e., in the time frame $\alpha$, $w_\alpha$ counts the number of eigenmodes localized around the entanglement cuts in the ES, instead of the number of zero or $\pi$ eigenmodes in the Floquet quasienergy spectrum. To the best of our knowledge, this is the first time that an explicit physical meaning is given to the winding numbers $(w_1,w_2)$ of symmetric-frame Floquet operators. It further bridges a gap between the theories of quasienergy-band topology and entanglement topology in Floquet systems.

Experimentally, our model may be realized in cold atom systems, where the engineering of periodically quenched \cite{FloQC1} and harmonically driven \cite{FloQC2} Kitaev chains were considered in previous studies. In cold atoms, the superconducting pairing term may be realized via p-wave Feshbach resonances \cite{Pwave1}, synthetic spin-orbit couplings \cite{Pwave2}, or the orbital degrees of freedom in association with s-wave interactions \cite{Pwave3}. The Floquet driving could be introduced through the periodic modulation of magnetic fields \cite{FloMod1}, laser fields \cite{FloMod2} or Raman-coupling amplitudes \cite{FloMod3}. The $\delta$-kicking may then be implemented via a short-pulsed modulation with a large amplitude. Replacing the $\delta$-kicking on the hopping or pairing amplitude with time-periodic quenching could yield the same Floquet superconducting phases upon suitable rescaling of system parameters. The topological winding numbers of our system may be extracted from the mean chiral displacements of initially localized wavepackets in each symmetric time frame \cite{MCDExp1,MCDExp2}. Finally, the Floquet Majorana zero and $\pi$ edge modes may be detected by applying spatially resolved rf spectroscopy and in situ imaging techniques \cite{FloQC1}. Note in passing that the ranges of system parameters for us to observe phases with large winding numbers and many Floquet Majorana modes are essentially available under current experimental conditions \cite{FloQC1,FloQC2,FloQC3}. Therefore, we believe that the realization of our models and the detection of their topological properties should be within reach in near-term experiments.

In future work, it
is interesting to check the transport properties \cite{Junc1} and the robustness of Floquet superconducting
phases found here to disorder \cite{Disord1} and many-body interactions. Possible
experimental probes of the ES, EE and entanglement winding numbers
in Floquet systems also deserve to be explored.

%\vspace{0.5cm}
\begin{acknowledgments}
	L.Z. acknowledges Raditya Weda Bomantara for helpful suggestions. This work is supported by the National Natural Science Foundation of China (Grant Nos.~12275260 and 11905211), the Fundamental Research Funds for the Central Universities (Grant No.~202364008), and the Young Talents Project of Ocean University of China.
\end{acknowledgments}

\appendix

\section{Floquet operator in the BdG representation}\label{sec:UBdG}

We can express the Floquet operators of both the PKKC1 and PKKC2 in the BdG basis of fermions in real space. This can be achieved by performing the Bogoliubov transformation in Eq.~(\ref{eq:BdG}). There the $\hat{f}_{j}^{\dagger}$ and $\hat{f}_{j}$ are creation and annihilation operators of normal fermions, satisfying the anticommutation relations $\{\hat{f}_{i},\hat{f}_{j}\}=\{\hat{f}_{i}^{\dagger},\hat{f}_{j}^{\dagger}\}=0$ and $\{\hat{f}_{i},\hat{f}_{j}^{\dagger}\}=\delta_{ij}$. The Hamiltonian $\hat{H}_{\rm I}(t)$ or $\hat{H}_{\rm II}(t)$ of the PKKC1 or PKKC2 is assumed to be diagonalized in the basis $\{\hat{f}_{1},...,\hat{f}_{L},\hat{f}_{1}^{\dagger},...,\hat{f}_{L}^{\dagger}\}$, so that it can be expressed as $\hat{H}_s(t)=\sum_{\ell}{\cal E}_{\ell}(t)\hat{f}_{\ell}^{\dagger}\hat{f}_{\ell}$ for $s={\rm I}$ or II, with ${\cal E}_{\ell}(t)$ being the instantaneous eigenvalue. Using the operator equality $[\hat{A}\hat{B},\hat{C}]=\hat{A}\{\hat{B},\hat{C}\}-\{\hat{A},\hat{C}\}\hat{B}$ and Eq.~(\ref{eq:BdG}), it is straightforward to show that $[\hat{H}_s(t),\hat{f}_{j}]=-{\cal E}_{j}(t)\hat{f}_{j}$, $[\hat{H}_s(t),\hat{f}_{j}^{\dagger}]={\cal E}_{j}(t)\hat{f}_{j}^{\dagger}$, and
\begin{equation}
	[\hat{H}_s(t),\hat{c}_{n}]=-\sum_{j}{\cal E}_{j}(t)(u_{jn}\hat{f}_{j}-v_{jn}^{*}\hat{f}_{j}^{\dagger}).\label{eq:HtCn1}
\end{equation}

Meanwhile, we can compute the commutator $[\hat{H}_s(t),\hat{c}_{n}]$ directly with the $\hat{H}_s(t)$ in Eq.~(\ref{eq:H1}) or (\ref{eq:H2}) and the $\hat{c}_{n}$ in Eq.~(\ref{eq:BdG}), yielding
\begin{alignat}{1}
	[\hat{H}_s(t),\hat{c}_{n}]= & -\mu\sum_{j}(u_{jn}\hat{f}_{j}+v_{jn}^{*}\hat{f}_{j}^{\dagger})-\frac{J(t)}{2}\sum_{j}[(u_{jn-1}+u_{jn+1})\hat{f}_{j}+(v_{jn-1}^{*}+v_{jn+1}^{*})\hat{f}_{j}^{\dagger}]\nonumber\\
	& -\frac{\Delta(t)}{2}\sum_{j}[(v_{jn-1}-v_{jn+1})\hat{f}_{j}+(u_{jn-1}^{*}-u_{jn+1}^{*})\hat{f}_{j}^{\dagger}],\label{eq:HtCn2}
\end{alignat}
where $\Delta(t)\equiv\Delta\delta(t)$ for $s={\rm I}$ and $J(t)\equiv J\delta(t)$ for $s={\rm II}$. For the Eqs.~(\ref{eq:HtCn1}) and (\ref{eq:HtCn2}) to be equal, the coefficients in front of $\hat{f}_{j}^{\dagger}$ and $\hat{f}_{j}$ must be separately identical. Therefore, after dropping the redundant band index $j$, we obtain the BdG self-consistent equations in the fermionic basis for the PKKC1 and PKKC2, as shown in Eqs.~(\ref{eq:BdG3})--(\ref{eq:BdG6}). These equations can be compactly written as $\mathsf{H}_s(t)|\psi\rangle={\cal E}(t)|\psi\rangle$, where
\begin{equation}
	\mathsf{H}_{\rm I}(t)=\frac{\Delta(t)}{2i}\begin{pmatrix}0 & 1 & 0 & 0 & \cdots & 0\\
		-1 & 0 & 1 & 0 & \cdots & 0\\
		0 & -1 & \ddots & \ddots & \ddots & \vdots\\
		0 & 0 & \ddots & \ddots & 1 & 0\\
		\vdots & \vdots & \ddots & -1 & 0 & 1\\
		0 & 0 & \cdots & 0 & -1 & 0
	\end{pmatrix}_{L\times L}\otimes\sigma_{y} + \begin{pmatrix}\mu & \frac{J}{2} & 0 & 0 & \cdots & 0\\
	\frac{J}{2} & \mu & \frac{J}{2} & 0 & \cdots & 0\\
	0 & \frac{J}{2} & \ddots & \ddots & \ddots & \vdots\\
	0 & 0 & \ddots & \ddots & \frac{J}{2} & 0\\
	\vdots & \vdots & \ddots & \frac{J}{2} & \mu & \frac{J}{2}\\
	0 & 0 & \cdots & 0 & \frac{J}{2} & \mu
	\end{pmatrix}_{L\times L}\otimes\sigma_{z}
\end{equation}
for the PKKC1,
\begin{equation}
	\mathsf{H}_{\rm II}(t)=\frac{\Delta}{2i}\begin{pmatrix}0 & 1 & 0 & 0 & \cdots & 0\\
		-1 & 0 & 1 & 0 & \cdots & 0\\
		0 & -1 & \ddots & \ddots & \ddots & \vdots\\
		0 & 0 & \ddots & \ddots & 1 & 0\\
		\vdots & \vdots & \ddots & -1 & 0 & 1\\
		0 & 0 & \cdots & 0 & -1 & 0
	\end{pmatrix}_{L\times L}\otimes\sigma_{y} + \begin{pmatrix}\mu & \frac{J(t)}{2} & 0 & 0 & \cdots & 0\\
		\frac{J(t)}{2} & \mu & \frac{J(t)}{2} & 0 & \cdots & 0\\
		0 & \frac{J(t)}{2} & \ddots & \ddots & \ddots & \vdots\\
		0 & 0 & \ddots & \ddots & \frac{J(t)}{2} & 0\\
		\vdots & \vdots & \ddots & \frac{J(t)}{2} & \mu & \frac{J(t)}{2}\\
		0 & 0 & \cdots & 0 & \frac{J(t)}{2} & \mu
	\end{pmatrix}_{L\times L}\otimes\sigma_{z},
\end{equation}
for the PKKC2, and $|\psi\rangle=(u_{1},v_{1},...,u_{n},v_{n},...,u_{L},v_{L})^{{\rm \top}}$. The related dynamical equation of the system now takes the form of $i\partial_t|\Psi\rangle=\mathsf{H}_s(t)|\Psi\rangle$. The Floquet matrix in Eq.~(\ref{eq:UBdG}) of the main text can then be obtained by integrating this equation over a driving period.

\section{Calculation details for the PKKC2}\label{sec:APP3}

The bulk Floquet spectrum of the PKKC2 is given by the solution of eigenvalue equation $U_{{\rm II}}(k)|\psi\rangle=e^{-iE}|\psi\rangle$
with the $U_{{\rm II}}(k)$ in Eq.~(\ref{eq:U2k}). The resulting
quasienergy dispersion reads 
\begin{equation}
	\pm E(k)=\pm\arccos\left[\cos(r_{1})\cos(r_{2})-\frac{\mu}{r_{2}}\sin(r_{1})\sin(r_{2})\right],\label{eq:E2k}
\end{equation}
where $r_{1}=J\cos k$ and $r_{2}=\sqrt{\Delta^{2}\sin^{2}k+\mu^{2}}$.
There are two Floquet bands symmetric with
respect to $E=0$ and $\pi$. They may touch with each other at $E=0$
($E=\pm\pi$) when $\cos[E(k)]=1$ ($\cos[E(k)]=-1$). Since $|\cos(r_{1})\cos(r_{2})-\frac{\mu}{r_{2}}\sin(r_{1})\sin(r_{2})|\leq|\cos(r_{1})\cos(r_{2})|+|\frac{\mu}{r_{2}}||\sin(r_{1})\sin(r_{2})|\leq1$,
the equality can be reached if either $|\mu/r_{2}|=1$ or $|\sin(r_{1})\sin(r_{2})|=0$.
In the former case, we must have $\Delta\sin k=0$, yielding the gapless
condition $\cos(J\pm\mu)=0$ for $k=0,\pi$. In the latter case, we
will have $\cos(r_{1})\cos(r_{2})=\pm1$ when the quasienergy gap
closes. Putting together, we find the equations that describing all
possible phase boundaries in the parameter space of the PKKC2, i.e.,
\begin{equation}
	J=p\pi\pm\mu\qquad p\in\mathbb{Z},\label{eq:PB21}
\end{equation}
or
\begin{equation}
	\frac{p^{2}\pi^{2}}{J^{2}}+\frac{q^{2}\pi^{2}-\mu^{2}}{\Delta^{2}}=1,\label{eq:PB22}
\end{equation}
for $p,q\in\mathbb{Z}$ assuming $|J|\geq|p\pi|$ and $\sqrt{\Delta^{2}+\mu^{2}}\geq|q\pi|$.
The union of Eqs.~(\ref{eq:PB21}) and (\ref{eq:PB22}) yields
the topological phase boundaries of the PKKC2 in the parameter space
$(J,\mu,\Delta)$.

The topological phase diagrams of the PKKC2 are obtained by first identifying
$h_{a}(k)=J\cos k\sigma_{z}$, $h_{b}(k)=\Delta\sin k\sigma_{y}+\mu\sigma_{z}$
in Eq.~(\ref{eq:Uk}) referring to the $U_{{\rm II}}(k)$ in Eq.~(\ref{eq:U2k}),
and then computing the winding numbers $(w_{0},w_{\pi})$ following
Eqs.~(\ref{eq:Uk})--(\ref{eq:w0p}).

The quasienergy gap functions 
{[}Eq.~(\ref{eq:G0P}){]} under the PBC can be obtained by inserting
the $E(k)$ in Eq.~(\ref{eq:E2k}) into the Eq.~(\ref{eq:G0P}). The
Floquet spectra under the OBC can be obtained by solving the eigenvalue equation $\mathsf{U}|\psi\rangle=e^{-iE}|\psi\rangle$
for the $\mathsf{U}$ of PKKC2 in Eq.~(\ref{eq:UBdG}) under the OBC,
where the matrices in the two exponentials of $\mathsf{U}$ are given
by $\mathsf{H}_{a}=\frac{1}{2}\sum_{n}(J|n\rangle\langle n+1|+{\rm H.c.})\otimes\sigma_{z}$
and $\mathsf{H}_{b}=\frac{1}{2}\sum_{n}(\mu|n\rangle\langle n|\otimes\sigma_{z}-i\Delta|n\rangle\langle n+1|\otimes\sigma_{y}+{\rm H.c.})$.

To obtain the ES, we first identify the $\hat{{\cal H}}_{a}$ and
$\hat{{\cal H}}_{b}$ of PKKC2 from Eqs.~(\ref{eq:HBdG2}) and
(\ref{eq:UESEE}) as
\begin{equation}
	\hat{{\cal H}}_{a}=\frac{1}{2}\sum_{n}(J\hat{\boldsymbol{{\bf c}}}_{n}^{\dagger}\sigma_{z}\hat{\boldsymbol{{\bf c}}}_{n+1}+{\rm H.c.}),\label{eq:Ha2}
\end{equation}
\begin{equation}
	\hat{{\cal H}}_{b}=\frac{1}{2}\sum_{n}(\mu\hat{\boldsymbol{{\bf c}}}_{n}^{\dagger}\sigma_{z}\hat{\boldsymbol{{\bf c}}}_{n}-i\Delta\hat{\boldsymbol{{\bf c}}}_{n}^{\dagger}\sigma_{y}\hat{\boldsymbol{{\bf c}}}_{n+1}+{\rm H.c.}).\label{eq:Hb2}
\end{equation}
Inserting them into the $\hat{{\cal U}}_{1}$ and $\hat{{\cal U}}_{2}$
in Eqs.~(\ref{eq:UI}) and (\ref{eq:UII}), we find the Floquet operators
of the system $\hat{{\cal H}}_{{\rm II}}(t)$ {[}Eqs.~(\ref{eq:HBdG2}){]}
in symmetric time frames $\alpha=1,2$. For a spatially bisected,
half-filled lattice under the PBC, we can follow the steps
in Sec.~\ref{sec:PKKC} to get the bipartite ES $\zeta_{\alpha}$ and
EE $S_{\alpha}$ of $\hat{{\cal U}}_{\alpha}$ in the symmetric time
frame $\alpha$. To obtain the
entanglement winding numbers $(W_{1},W_{2})$, we plug the Eqs.~(\ref{eq:Ha2})
and (\ref{eq:Hb2}) into Eq.~(\ref{eq:UESEE}) and go through the
calculations from Eqs.~(\ref{eq:UI}) to (\ref{eq:Qa}).
A complete topological characterization of the PKKC2
can then be achieved by combining the entanglement information obtained
in the two time frames.

\section{Derivation of Eq.~(\ref{eq:N12n0p})}\label{sec:App2}
Here we present further details for the derivation of Eq.~(\ref{eq:N12n0p}) from Eqs.~(\ref{eq:BBC1})--(\ref{eq:EBBC1}) in the main text. We will do this by considering all possible situations.

(i) $n_{0}>n_{\pi}$: In this case, according to Eq.~(\ref{eq:EBBC1}), we have
$|W_{1}+W_{2}|>|W_{1}-W_{2}|$, which implies that $W_{1}$ and $W_{2}$
have the same signs. We can thus write $|W_{1}+W_{2}|=|W_{1}|+|W_{2}|$
and $|W_{1}-W_{2}|=||W_{1}|-|W_{2}||$. According to Eq.~(\ref{eq:BBC1}), we
further have $|W_{1}|+|W_{2}|=\frac{1}{2}(N_{1}+N_{2})$ and $||W_{1}|-|W_{2}||=\frac{1}{2}|N_{1}-N_{2}|$.
Therefore, we find the relations
\begin{equation}
	n_{0}+n_{\pi}=\frac{1}{2}(N_{1}+N_{2}+|N_{1}-N_{2}|)=\max(N_{1},N_{2}),\label{eq:AppB1}
\end{equation}
\begin{equation}
	|n_{0}-n_{\pi}|=\frac{1}{2}|N_{1}+N_{2}-|N_{1}-N_{2}||=\min(N_{1},N_{2}).\label{eq:AppB2}
\end{equation}

(ii) $n_{0}<n_{\pi}$: In this case, according to Eq.~(\ref{eq:EBBC1}), we have
$|W_{1}+W_{2}|<|W_{1}-W_{2}|$, which implies that $W_{1}$ and $W_{2}$
have the opposite signs. We can thus write $|W_{1}+W_{2}|=||W_{1}|-|W_{2}||$
and $|W_{1}-W_{2}|=|W_{1}|+|W_{2}|$. According to Eq.~(\ref{eq:BBC1}), we further
have $|W_{1}|+|W_{2}|=\frac{1}{2}(N_{1}+N_{2})$ and $||W_{1}|-|W_{2}||=\frac{1}{2}|N_{1}-N_{2}|$.
Therefore, we find the relations
\begin{equation}
	n_{0}+n_{\pi}=\frac{1}{2}(|N_{1}-N_{2}|+N_{1}+N_{2})=\max(N_{1},N_{2}),\label{eq:AppB3}
\end{equation}
\begin{equation}
	|n_{0}-n_{\pi}|=\frac{1}{2}||N_{1}-N_{2}|-N_{1}-N_{2}|=\min(N_{1},N_{2}).\label{eq:AppB4}
\end{equation}

(iii) $n_{0}=n_{\pi}$: In this case, according to Eq.~(\ref{eq:EBBC1}), we have
$|W_{1}+W_{2}|=|W_{1}-W_{2}|$, which implies that either $W_{1}=0$
or $W_{2}=0$. In each situation, we have $|W_{1}+W_{2}|=|W_{1}|+|W_{2}|$
and $|W_{1}-W_{2}|=||W_{1}|-|W_{2}||$, such that $|W_{1}|+|W_{2}|=\frac{1}{2}(N_{1}+N_{2})$
and $||W_{1}|-|W_{2}||=\frac{1}{2}|N_{1}-N_{2}|$ due to Eq.~(\ref{eq:BBC1}).
Therefore, we again arrive at the Eqs.~(\ref{eq:AppB1})--(\ref{eq:AppB2}).
The Eq.~(\ref{eq:N12n0p}) is thus coincident with the predictions of  Eqs.~(\ref{eq:BBC1})--(\ref{eq:EBBC1}).

\end{document}